\documentclass[%
reprint,
 amsmath,amssymb,
 aps,
floatfix,
]{revtex4-2}
\usepackage{graphicx}
\usepackage{dcolumn}
\usepackage{bm}
\usepackage{color}
\usepackage{xcolor}
\usepackage{soul}
\usepackage{braket}

\begin{document}
\preprint{APS/123-QED}
\title{Experimental Confirmation of First-Principles Thermal Conductivity in Zirconium-Doped ThO$_2$}
\author{Ella Kartika Pek$^1$}
\author{Zilong Hua$^1$}
\author{Amey Khanolkar$^1$}
\author{J. Matthew Mann$^2$}
\author{David B. Turner$^3$}
\author{Karl Rickert$^4$}
\author{Timothy A. Prusnick$^4$}
\author{Marat Khafizov$^5$}
\author{David H. Hurley$^1$}
\author{Linu Malakkal$^6$}
\email[]{linu.malakkal@inl.gov}

\affiliation{$^1$Condensed Matter and Materials Physics Group, Idaho National Laboratory, Idaho Falls, ID, 83415, USA}
\affiliation{$^2$Air Force Research Laboratory, Sensors Directorate, 2241 Avionics Circle, WPAFB, OH, 45433, USA}
\affiliation{$^3$Azimuth Corporation, Fairborn, Ohio 45324, USA}
\affiliation{$^4$KBR, 2601 Mission Point Boulevard, Suite 300, Dayton, Ohio 45431,USA}
\affiliation{$^5$Department of Mechanical and Aerospace Engineering, The Ohio State University, 201 West 19th Ave, Columbus, Ohio 43210, United States}
\affiliation{$^6$Computational Mechanics and Materials Department, Idaho National Laboratory, Idaho Falls, Idaho 83415, USA}
\date{\today}

\begin{abstract}

The degradation of thermal conductivity in advanced nuclear fuels due to the accumulation of fission products and irradiation-induced defects is inevitable, and must be considered as part of safety and efficiency analyses of nuclear reactors. This study examines the thermal conductivity of a zirconium-doped ThO$_2$ crystal, synthesized via the hydrothermal method using a spatial domain thermo-reflectance technique. Zirconium is one of the soluble fission products in oxide fuels that can effectively scatter heat-carrying phonons in the crystalline lattice of fuel. Thus, thermal property measurements of zirconium-doped ThO$_2$ single crystals provide insights into the effects of substitutional zirconium doping, isolated from extrinsic factors such as grain boundary
scattering. The experimental results are compared with first-principles calculations of the lattice thermal conductivity of ThO$_2$, employing an iterative solution of the Peierls-Boltzmann transport equation. Additionally, the non-perturbative Green’s function methodology is utilized to compute phonon-point defect scattering rates, accounting for local distortions around point defects, including mass difference changes, interatomic force constants, and structural relaxation. The congruence between the predicted results from first-principles calculations and the measured temperature-dependent thermal conductivity validates the computational methodology. Furthermore, the methodologies employed in this study enable systematic investigations of thermal conductivity reduction by
fission products, potentially leading to the development of more accurate fuel performance codes.

\end{abstract}

\maketitle


\section{\label{sec:level1}Introduction \protect}
Thorium dioxide (ThO$_2$) is emerging as a pivotal material for advanced thorium based nuclear fuel cycles, and is anticipated to be able to meet future energy demands while offering a higher safety margin alternative to uranium dioxide, which is currently the predominant fuel for uranium based fuel cycles \cite{Herring2001,Hania2012304T,Ashley_2012,KUTTY_2008,hurley2022thermal,iaea2005,DESKINS2022,MALAKKAL2019507}. Increasing demand for ThO$_2$ is further underscored by its adoption by nuclear industries \cite{Clean_energy,IREL_indian2020}. Among the various chemical and physical properties of nuclear fuel, thermal conductivity is a critical factor that influences the efficiency and safety of nuclear reactors\cite{Olander_1976}. Therefore, understanding the thermal transport properties of ThO$_2$ under different reactor conditions is essential \cite{hurley2022thermal}. During reactor irradiation, energetic fission fragments introduce a range of lattice defects, including point defects, small defect clusters, and extended defects such as dislocation loops, inert gas bubbles, grain subdivision-induced grain boundaries, and defect segregation at grain boundaries \cite{Olander_1976,Kawano2023}. These defects scatter phonons leading to a degradation in thermal conductivity \cite{hurley2022thermal}. Existing fuel performance codes \cite{Bison,Transuranus_MAGNI_2021,ALCYONE_INTROINI_2024,VANUFFELEN_2019_review} account for this degradation using thermal conductivity correlations that have multiplicative factors that each corrects for dissolved and precipitated fission products, porosity, and empirical corrections for irradiation-induced damage \cite{LUCUTA_1996,FERRIGNO_2024,MAGNI2020152410}. While these empirical correlations are valuable, they come with inherent limitations. Their application is restricted to the conditions under which they were developed. Moreover, these correlations do not provide mechanistic insights, oversimplify complex phenomena, and lack the ability to predict novel behaviors. To overcome these limitations, it is essential to complement empirical correlations with mechanistic models and continuously validate them with experimental data. Several report have aimed to provide a more mechanistic description of conductivity reduction due to irradiation--induced defects \cite{FERRIGNO_2023} inspired by a fundamental understanding of thermal transport phenomena [\cite{hurley2022thermal,DENNETT2021,Liu_2016_MD, JIN_2022_MD_tho}. The impact of dissolved fission products inspired by classical theory for phonon mediated thermal transport \cite{Klemens_1955}, is captured using empirical values for phonon scattering cross-sections \cite{LUCUTA_1996}. Some efforts have been made to further refine these scattering cross-sections based on experimental measurements \cite{BONEV_2023_JNM,HORII_2024_JNM}, but there have been very limited efforts to improve mechanistic understanding for these cross-sections from first principles \cite{Liu_2016_MD, JIN_2022_MD_tho, CHEN_2024}.

While the reduction in thermal conductivity in nuclear fuels results from the combined effects of various defects generated during the fission process,  it is crucial to systematically analyze the impact of each defect type individually for a comprehensive understanding. Among the various defect types, we focus on examining the role of irradiation-induced small-scale point defects, which notably influence thermal conductivity during the initial stages of damage accumulation  \cite{Khafizov_2016_investigation,Dennett_2020,DESKINS2022}. The irradiation-induced point defects mainly include vacancies, interstitials, substitutional defects, Frenkel pairs, and Schottky defects. Grimes \text{et al.} \cite{Grimes_1991} have shown that the inert gas fission products, such as xenon (Xe) and krypton (Kr), in uranium dioxide are estimated to be 15$\%$ of the total fission yield. Other significant fission products include zirconium (Zr) and iodine (I) \cite{RONCHI_2004}. Despite the large number of fission products present in nuclear fuels, the effect of substitutional fission products on the thermal transport of ThO$_2$ is still not well understood. Therefore, in this work, we use Zr-doped ThO$_2$  to understand how fission products reduce ThO$_2$'s thermal conductivity, combining first-principles calculations and experimental analysis.

Numerous studies have investigated the effects of various defects, fission products, and uranium doping on the thermal conductivity of thoria \cite{DENNETT2021,DESKINS2022,Dennett_2020,HUA_2024709,Malakkal_PRM,Masato_2024,SAOUDI_2018,TURNBULL_2023}. For example, Cooper \text{et al.} \cite{COOPER201529} observed a reduction in thermal conductivity due to an irregular cation lattice in (U$_x$Th$_{1-x}$)O$_2$ solid solutions, using the non-equilibrium molecular dynamics method (NEMD), from 300 to 2000 K. Hua et al. \cite{HUA_2024709} recently reported thermal conductivity suppression in U-doped ThO$_2$ due to phonon-spin interactions. Park \text{et al.} \cite{PARK2018198} utilized reverse non-equilibrium molecular dynamics (r-NEMD) to examine the influence of vacancies and uranium substitutional defects on ThO$_2$ thermal transport. Rahman \text{et al.} \cite{RAHMAN2020152050} explored how fission-generated products (xenon and krypton) and vacancies affect ThO$_2$’s thermal conductivity within the 300-1500 K range using NEMD simulations. Notably, their study did not address the influence of Zr fission product generation on ThO$_2$’s thermal conductivity. Recently, Jin \text{et al.} \cite{JIN2022Impact} used NEMD simulations to determine the phonon-defect scattering cross section for various defects, which can be applied to reduced-order analytical models. Chen et al. \cite{CHEN_2024} \text investigated how point defects in ThO$_2$ impact phonon mode-resolved thermal transport and predicted a reduction in thermal conductivity due to vacancies and interstitials. Besides MD simulations, Deskins et al.\cite{Deskins_2021} examined the impact of point defects on ThO$_2$’s thermal conductivity using the Klemens model for phonon relaxation times, which accounts for changes in mass and induced lattice strain associated with point defects. These studies generally overlook quantum effects by assuming that the phonon energy distribution is described using Boltzmann statistics rather than the more accurate Planck distribution \cite{Zhou_2018}. Additionally, the accuracy of MD simulation results depends on the precision of the interatomic potential. Jin \text{et al.} \cite{Jin_2021}  indicated that the empirical potential used in previous studies requires further optimization for accurately predicting ThO$_2$’s thermal conductivity in both perfect crystals and those with complex defects. To overcome these limitations, Malakkal et al. \cite{Malakkal_PRM} predicted phonon scattering due to point defects and their impact on the thermal conductivity of ThO$_2$ using an \textit {ab initio} calculation utilizing the Green-function T-matrix method. In that study, the thermal conductivity degradation of ThO$_2$ due to a selected set of point defects, including the substitutions of transmutation and radioactive decay product helium (He$_{Th}$) and fission products, such as  krypton (Kr$_{Th}$), zirconium (Zr$_{Th}$), iodine (I$_{Th}$), and xenon (Xe$_{Th}$) in thorium sites, along with other point defects, such as the single vacancy of thorium (V$_{Th}$) and oxygen (V$_{O}$) and three configurations of Schottky defects (SD$_{100}$, SD$_{110}$, SD$_{111}$), were carried out. Despite several theoretical studies, a direct experimental comparison of the predicted impact of point defects on the thermal conductivity of ThO$_2$ has been lacking. Therefore, in this work, we address this gap by validating the first principles calculation of the reduction in thermal conductivity due to substitutional point defects by measuring the thermal conductivity of zirconium-doped ThO$_2$.

To accurately represent the impact of specific point defects on ThO$_2$, it is crucial to work with single crystals of ThO$_2$. Otherwise, grain boundary scattering in polycrystalline materials can obscure the influence of fission product defects on the thermal conductivity. Therefore, single crystal ThO$_2$ is essential to isolate the effects caused by Zr$_{Th}$ substitutional defects. In this study, we synthesized single crystals using the hydrothermal growth method \cite{Mann_2010}. This method has been proven to produce high-quality single crystal thoria, as evidenced by its high thermal conductivity \cite{Mann_2010,HUA_2024709}. To mimic the process of fission fragment (Zr) deposition onto nuclear fuel, we intentionally doped the ThO$_2$ single crystal with one atomic percent of Zr. The thermal conductivity of Zr-doped ThO$_2$ was measured using the spatial domain thermoreflectance (SDTR) method, which offers improved accuracy as it does not require prior knowledge of the beam spot size. Both calculations and experiments focused on the low-temperature range, where three-phonon scattering is less dominant. By measuring thermal conductivity below room temperature, we can isolate the effects of Zr substitutional defects on the thermal conductivity of ThO$_2$. This approach allows us to validate the accuracy of theoretical calculations.

\section{\label{sec:level1}Experimental and computational details \protect}
\subsection{Synthesis of samples}
\begin{figure}[h!]
\includegraphics [width=0.9\columnwidth]{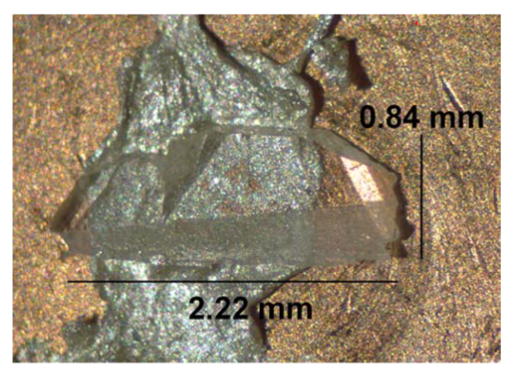}
\caption{\label{fig:fig1} The synthesized crystal of Zr-doped-ThO$_2$ using the hydrothermal method mounted on copper holders with silver paste.}
\end{figure}

Single crystals of zirconium-doped ThO$_2$ (Figure.~\ref{fig:fig1}) and undoped ThO$_2$ were synthesized using supercritical hydrothermal solutions. A powdered feedstock consisting of 21.93 mg of zirconium oxide (Alfa Aesar, 99+$\%$) and 2,383.41 mg of ThO$_2$ (IBI Labs, 99.99$\%$) was placed into a silver tube with a diameter of 3/8 in and a length of 8 in. 4.8 mL of 6M cesium fluoride (Alfa Aesar, 99.99$\%$) were added as the mineralizer solution.  The tube was then sealed and placed into a 250 mL Inconel autoclave containing 140 mL of counterpressure water. The bottom section of the autoclave, containing the powdered feedstock, was heated to 650°C, while the upper section, corresponding to the crystallization zone, was maintained at 600°C. Pressure within the autoclave reached 19 kpsi. These pressure and temperature conditions were maintained for 45 days, followed by a 24-hour cooldown to room temperature. The silver tube was retrieved from the autoclave and opened to obtain the zirconium doped ThO$_2$ crystal. The undoped ThO$_2$ crystal was grown under nearly identical conditions, with only a minor change in the pressure at 24 kpsi. The synthesized samples were then characterized for structural and thermal properties. 

\subsection{Structural, luminescence, and thermal property measurements}
A Bruker M4 Tornado X-ray Fluorescence (XRF) system was employed to analyze the zirconium concentration in the ThO$_2$ sample. Thirty spots were randomly selected across the surface of the Zr-doped-ThO$_2$ sample and analyzed using a rhodium X-ray source (50 kV, 300 A) with a 120-second exposure. The X-ray beam for each spot was 15.7 $\mu$m in diameter. The average zirconium concentration was determined to be 1.13 at$\%$ with a standard deviation of 0.07. Raman measurements were performed using a Renishaw InVia Raman microscope system in
back scattering geometry to confirm the fluorite crystal structure and interrogate any additional vibrational modes due to zirconium substitution. Measurements were taken using a 532 nm laser. Scans were taken with 1800 l/mm gratings, centered at 1000 cm$^{-1}$ to cover the desired range of interest. The lasers were
focused onto the samples through a 50× long working distance objective with a 0.5 numerical
aperture. Forty scans were averaged at each point to maximize signal to noise ratios. Photoluminescence (PL) measurements were carried out on the same InVia system using a 325 nm HeCd laser for excitation in combination with a 150 l/mm grating to probe the electronic structure. The grating was centered at
650 nm to cover the effective range of the detector. In this case, the laser was focused onto the sample using a 15× near-ultraviolet objective with a 0.31 numerical aperture. Twenty scans per spot were averaged to achieve the desired signal to noise ratios.

The thermal conductivity and thermal diffusivity of pristine ThO$_2$ and Zr-doped ThO$_2$ were measured using the SDTR technique. SDTR is a pump-probe system utilizing continuous-wave lasers, with the pump being amplitude-modulated. One advantage of this measurement method is its insensitivity to optical spot size \cite{Feser_2012}, which improves measurement reliability and reproducibility. More detailed information about this measurement technique and the 3D thermal wave model used to extract the thermal conductivity and thermal diffusivity from the data can be found in \cite{Hurley_2015,Hua_2012,Maznev_1995}. For our SDTR setup, we used 660 nm (Coherent OBIS 660 nm) and 532 nm (Coherent Verdi 532 nm) continuous-wave lasers for the pump and probe, respectively. The power used was 3 mW for the pump and 0.3 mW for the probe. The pump laser primarily acts as the heating laser, introducing a temperature gradient on the sample’s surface. The probe laser detects the phase lag caused by the heating laser via thermoreflectance, which is collected using a photodetector. The photodetector is connected to a lock-in amplifier that is coupled to the modulation frequency of the pump laser. The laser beams are focused using a 50× long-working-distance objective lens, resulting in a 1 µm spot size with power of 3 mW and 0.3 mW for the pump and probe lasers, respectively. The thermal conductivity measurement was conducted at cryogenic temperatures ranging from 77 K to 300 K, with increments of 25 K (between 100 K-275 K).
The sample was mounted inside a liquid nitrogen-cooled optical cryostat (Cryo Industries model XEM), with the temperature maintained at the intended Zr-doped ThO$_{2}$ temperature, exhibiting a maximum fluctuation of 0.5 K. The pressure within the cryostat chamber was kept below 5 mTorr, and the chamber was initially purged with ultrahigh-purity nitrogen to prevent ice condensation on the sample surface. At each temperature, a minimum of four sets of measurements were conducted with three different pump modulation frequencies: 20, 50, and 100 kHz. For each frequency measurement, the pump was scanned across the probe beam over a 10 $\mu$m distance in each direction. Multiple data sets were collected to reduce the statistical uncertainty of the thermal conductivity measurements. 

To ensure strong optical absorption and a robust thermoreflectance signal, the sample was coated with 35 nm of gold using DC magnetron sputtering. Gold was selected as the transducer layer due to its high thermoreflectance coefficient at the probe wavelength (532 nm) \cite{Wilson_2012}. The thickness and thermal conductivity of the gold film at room temperature were determined by measuring a gold-coated reference glass (BK7) sample, which was coated alongside the Zr-doped ThO$_{2}$ sample at the same height. The laser intensity before and after transmitting BK7 with film was measured, and using the Beer-Lambert law, the thickness of gold film can be extracted from the attenuation of optical transmission. This method was validated by comparing the results to the ones from the picosecond acoustic measurements, and the difference is below the experimental uncertainty. To extract the thermal conductivity values, we input the known parameters into the data analysis, such as the thermal conductivity and heat capacity of gold, alongside with the heat capacity of the sample (ThO$_{2}$/ Zr-doped ThO$_{2}$). For both pristine ThO$_{2}$ and Zr-doped ThO$_{2}$, the density and heat capacity of pure ThO$_{2}$ were employed as referenced in \cite{Dennett_2020}. The thermal conductivity and heat capacity of gold at various temperatures were derived from prior low-temperature measurements of gold thin films and the heat capacity data provided in the referenced literature \cite{Dennett_2020}.

\subsection{Simulation details}
All density functional theory \cite{Kohn_1965} calculations in this study were conducted using the projector-augmented-wave \cite{blochl_1994} method within the Vienna Ab initio Simulation Package (VASP) \cite{Kresse_1996}, employing the local density approximation (LDA) \cite{Perdew_1981} pseudopotential for exchange and correlation. Notably, no Hubbard U correction \cite{Dudarev_1988_HubbardU} was applied, as previous findings indicated it did not enhance the property predictions of ThO$_2$ \cite{SHIELDS201699,Lu_2012}. Geometry optimization of ThO$_{2}$ (space group $Fm\bar{3}m$) was performed on a primitive unit cell by minimizing total energy with respect to cell parameters and atom positions using the conjugate gradient method. Energy convergence for ThO$_{2}$ was achieved using an electron wave vector grid and plane wave energy cutoff of a 12 × 12 × 12 mesh and 550 eV, respectively, with electronic energy convergence criteria set at $10^{-8}$ eV. The relaxed lattice parameter of ThO$_{2}$ at 0 K was found to be 5.529 Å, consistent with previously reported values \cite{Jin_2021} for the LDA functional from VASP and comparable to the experimental lattice constant of 5.60 Å \cite{Jin_2021}. Although the generalized gradient approximation functional predicted the lattice constant more accurately than the LDA, it significantly underestimated thermal conductivity, leading to the use of the LDA functional in this work. A 5 × 5 × 5 supercell with 375 atoms and k-points of 2 × 2 × 2 was used to evaluate harmonic force constants via the finite displacement method as implemented in PHONOPY \cite{TOGO2015}. Further details of the phonon dispersion spectra used in this work are available in reference \cite{Malakkal_PRM}. 

The third-order force constants (anharmonic force constants) were calculated using a  5 × 5 × 5 supercell at the gamma point using Thirdorder.py \cite{shengBTE_2014}, setting the force cutoff distance to the fifth nearest neighboring atoms. Given that ThO$_{2}$ is a polar material, the non-analytical contribution was considered. The Born charges and dielectric constant required for evaluating the non-analytical correction \cite{Wang2016} were calculated using density functional perturbation theory \cite{Baroni_2001_Phonons}. Phonon-phonon scattering processes were evaluated using Fermi's golden rule from the cubic force constants. The lattice thermal conductivity (\textit{$k_{L}$}) was calculated using the iterative solutions of the Boltzmann transport equation (BTE) as implemented in ShengBTE \cite{shengBTE_2014}. All \textit{$k_{L}$} values presented in this work are fully iterative solutions of the Peierls-Boltzmann equation. Converged \textit{$k_{L}$} values were obtained using a cubic force constant, considering the fifth nearest-neighbor interaction. The number of grid planes along each axis in the reciprocal space for solving the BTE was 24 × 24 × 24. The phonon scattering due to isotopic atoms was also included in the thermal conductivity calculation using the Tamura expression \cite{Tamura_1983}. The Zr-doped ThO$_{2}$ was created in a 5 × 5 × 5 supercell by replacing a thorium atom with a Zirconium atom. The defected supercell was relaxed while keeping the cell volume constant. A slight displacement of the defect atom or its nearest neighbors was introduced before relaxation to avoid the system getting trapped in a saddle point of the potential energy surface. The dynamic stability of the relaxed structures was confirmed by ensuring that the phonon band structure did not exhibit any modes with imaginary frequency. The interatomic force constants (IFCs) of Zr-doped ThO$_{2}$ were calculated using the finite-displacement method. Finally, the phonon-defect scattering rates were calculated on a uniformly spaced 24 × 24 × 24 q-point mesh with an 18 × 18 × 18 grid for the Green’s function methodology, as implemented by Katre et al. \cite{Ankita_2017}. More details of the calculations on the Zr-doped ThO$_{2}$ can be found in reference \cite{Malakkal_PRM}.

\section{\label{sec:level1}Results\protect}
The pristine ThO$_{2}$ and Zr-doped ThO$_{2}$ were synthesized using supercritical hydrothermal solutions. To mimic the substitutional defects that occur during the neutron irradiation of nuclear fuels in the fission process, we intentionally doped the ThO$_{2}$ crystal with one atomic percent of Zr. The Zr-doped ThO$_{2}$ crystal was grown to a size of 2.2 × 0.8 mm, as shown in Fig.~\ref{fig:fig1}. The grown crystal was oriented along the [111] axis. The dopant concentration and crystal quality were verified using XRF, Raman spectroscopy, and PL measurements. The composition of the Zr-doped ThO$_{2}$ was confirmed via a 30-point scan XRF to be 1.13 at$\%$. The XRF measurements did not detect any other impurities in the Zr-doped ThO$_{2}$ due to its resolution limit when the atomic percentage is below 0.10. 

\begin{figure}[h]
\includegraphics [width=0.95\columnwidth]{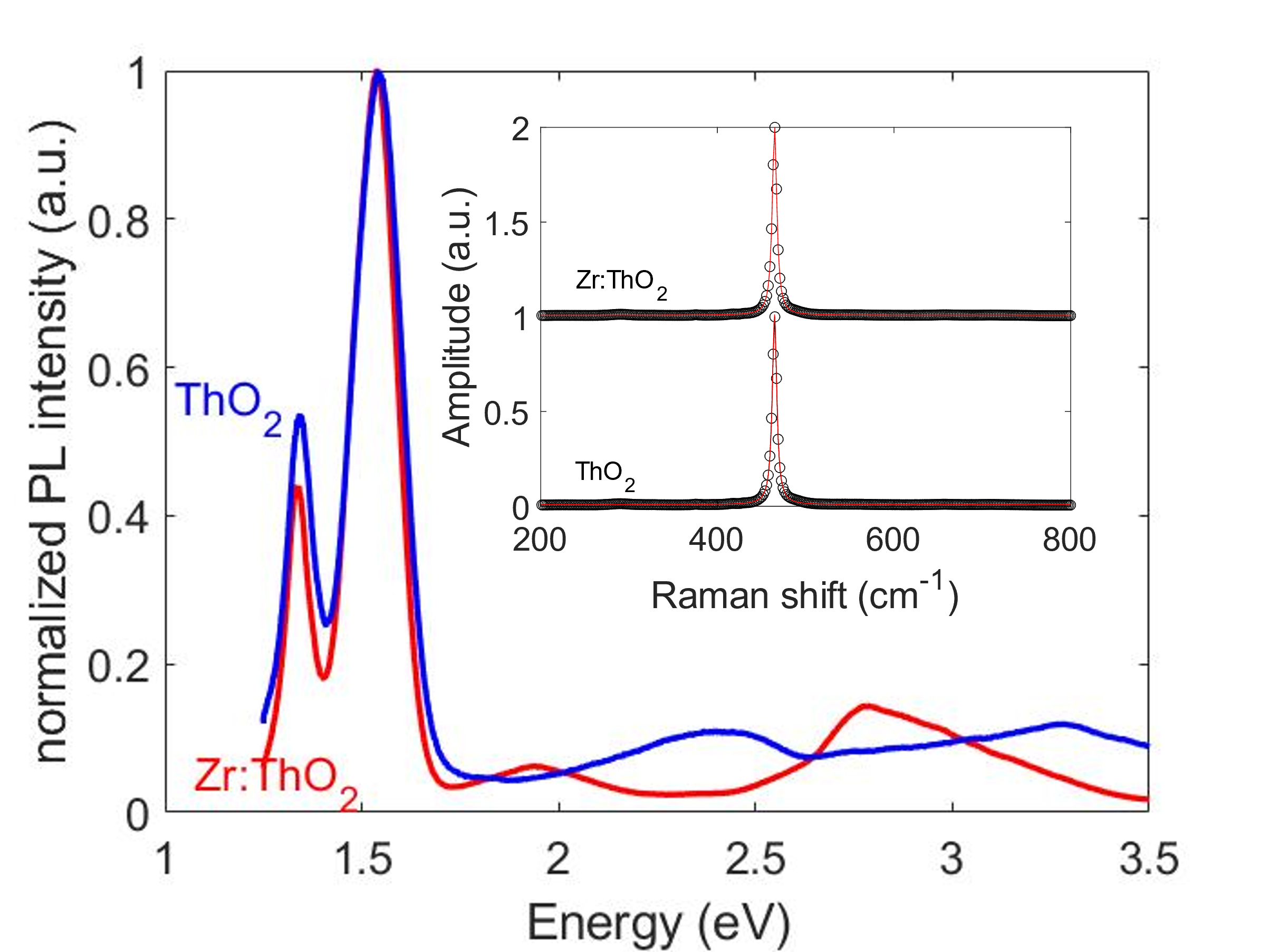}
\caption{\label{fig:fig2} The PL measurements ($\lambda$ = 325 nm) of pristine ThO$_{2}$ (blue) and Zr-doped ThO$_{2}$ (red). Raman measurement ($\lambda$ = 532 nm) of pristine ThO$_{2}$ (bottom) and Zr-doped ThO$_{2}$ (top) shown as inset figure.}
\end{figure}

Raman spectroscopy was conducted to assess the structural integrity and phase consistency of both pristine ThO$_{2}$ and Zr-doped ThO$_{2}$. The Raman spectra are presented in the inset of Figure \ref{fig:fig2}, with the pristine ThO$_{2}$ spectrum displayed at the bottom and the Zr-doped ThO$_{2}$ spectrum at the top. The peak centers for both samples remained consistent. Specifically, the peak center for pristine ThO$_{2}$ was observed at 465.7 cm$^{-1}$, while the Zr-doped ThO$_{2}$ exhibited a peak center at 464.7 cm$^{-1}$.  Moreover, the full width at half maximum (FWHM) values were nearly identical for both samples, with 6.04 cm$^{-1}$ for pristine ThO$_{2}$ and 6.05 cm$^{-1}$ for Zr-doped ThO$_{2}$. The PL measurements ($\lambda$ = 325 nm) were conducted on both pristine (blue) and Zr-doped ThO$_{2}$ (red) to investigate the impact of Zr doping on the electronic and optical properties of ThO$_{2}$. The PL spectra, depicted in Figure \ref{fig:fig2}, does not reveal notable differences between the pristine and Zr-doped ThO$_{2}$. Both PL spectra of pristine ThO$_{2}$ and Zr-doped ThO$_{2}$ exhibit prominent peaks at approximately 1.34 eV and 1.54 eV. 

\begin{figure}[h]
\includegraphics [width=0.95\columnwidth]{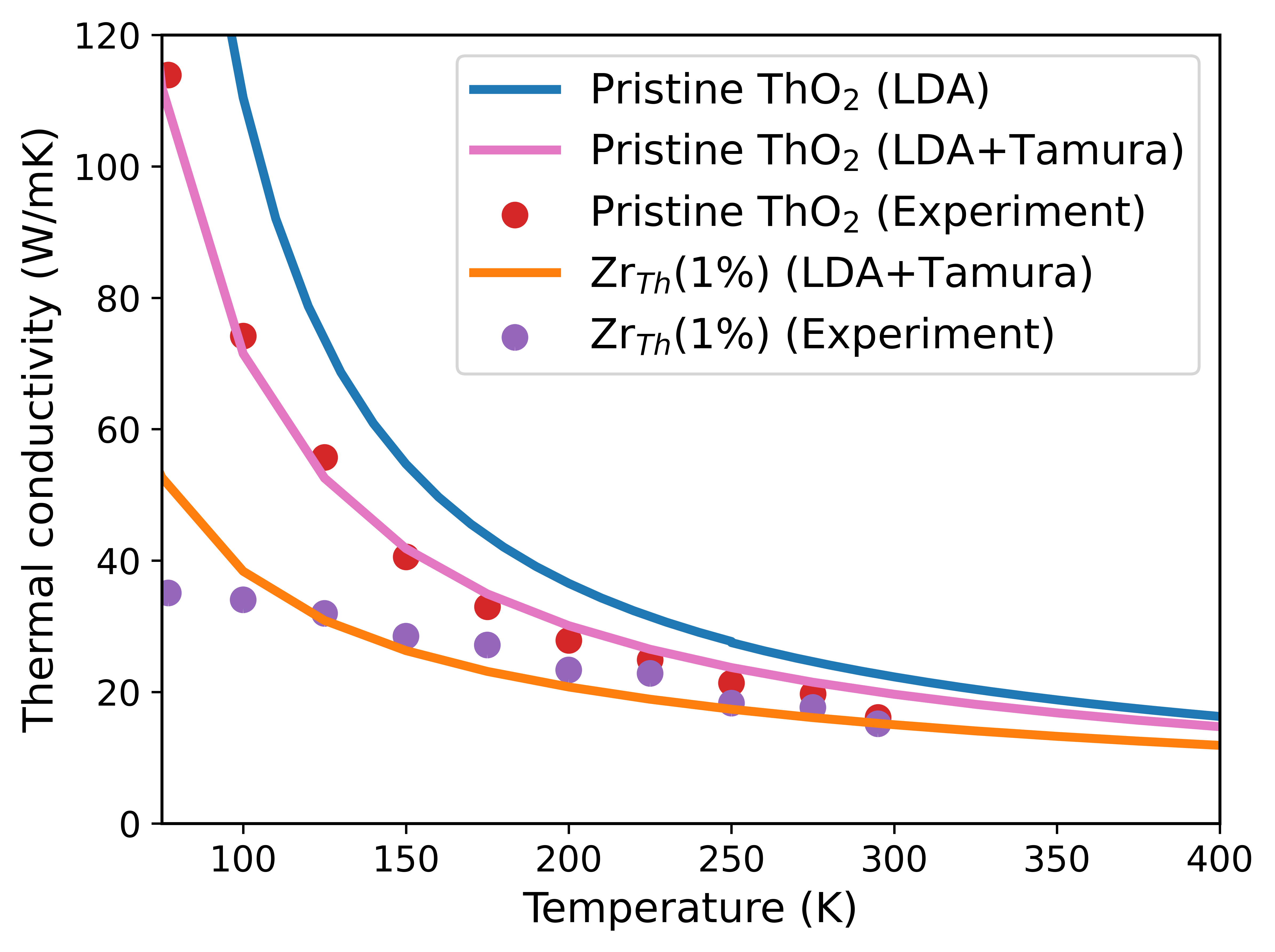}
\caption{\label{fig:epsart} This figure shows the comparison of the thermal conductivity of pristine and 1 at$\%$ Zr-doped ThO$_{2}$ from Green’s method computational results (solid lines) and the experimental results (data points). The measured thermal conductivity of the Zr-doped ThO$_{2}$ agrees well down to 100 K. The measured thermal conductivity of pristine ThO$_{2}$ agrees well with the corrected LDA+Tamura calculation.}
\end{figure}

Figure \ref{fig:epsart} provides a comparative analysis of the calculated and experimental thermal conductivity of ThO$_{2}$ and Zr-doped ThO$_{2}$ over the temperature range of 77–300 K. Multiple crystals were prepared for the pristine ThO$_{2}$ measurements, which exhibited slight variations in thermal conductivity due to difference in the amount of trace impurities in each sample. The thermal conductivity values reported in this study were measured using the same single crystal from our previous work \cite{HUA_2024709}, which showed the highest measured thermal conductivity (red solid dots). This suggests that the sample had minimal impurities and defects, making it ideal for comparison with theoretically predicted thermal conductivity for a defect-free crystal structure. Theoretical predictions for the lattice thermal conductivity of pristine ThO$_{2}$ were computed using the LDA functional (blue solid line). As is typical, the LDA functional tends to overestimate bond stiffness, and consequently overestimates thermal conductivity, representing an upper limit for theoretical thermal conductivity of ThO$_{2}$. The predicted lattice thermal conductivity, when accounting solely for three-phonon interactions, overestimated the values observed in experimental measurements. We acknowledge that the measured samples contain impurities, which we modeled using an isotopic scattering model, assuming these impurities are primarily substitutional. By incorporating isotope-phonon scattering using Tamura’s methodology, we achieved excellent agreement with the experimentally measured thermal conductivity. The results for pristine ThO$_{2}$ were consistent with the previous theoretical prediction \cite{Malakkal_PRM,Enda_2022_prb,MALAKKAL2019507}.  

\section{\label{sec:level1}Discussion\protect}
From the Raman measurements of pristine ThO$_{2}$ and Zr-doped ThO$_{2}$, we observed a negligible shift in the peak center. This indicates that substitutional doping with at$\%$ Zr does not significantly alter the Raman active mode T$_{2g}$, which is associated with the fluorite crystal structure \cite{Keramidas_1973}. These observations confirm that the Zr doping at 1 at$\%$ Zr does not induce significant structural distortions in the ThO$_{2}$ lattice, maintaining the material's structural integrity and crystalline quality. This is also in agreement with theoretical findings that the force constant perturbations are minimal when Zr occupies thorium sites.  \cite{Malakkal_PRM}. Furthermore, it is noteworthy that, in contrast to previously reported work on ion-irradiated ThO$_{2}$ single crystals \cite{RICKERT_2022}, which exhibited peaks corresponding to defects, our current study of Zr-doped ThO$_{2}$ shows only the T$_{2g}$ peak. This suggests that the Zr-doped samples exhibit minimal defect formation, likely because Zr atoms are effectively substituting for Th atoms in the lattice, which prevents the creation of vacancies or interstitials, thus maintaining the structural integrity of the material. The PL spectra of pristine ThO$_{2}$ exhibit prominent peaks at approximately 1.34 eV and 1.54 eV. These peaks align with previously reported values for as-grown ThO$_{2}$ single crystals under similar excitation  conditions \cite{RICKERT_2022}. Similarly, the PL spectra acquired on the Zr-doped ThO$_{2}$ crystal also exhibit peaks at approximately 1.34 eV and 1.54 eV. However, it is well known that the fundamental band-to-band gap energy parameter for ThO$_{2}$ is 5.4 eV \cite{Mock_2019} and there shouldn't be any peak below 5.4 eV in the case of pristine ThO$_{2}$. Therefore, these peaks are attributed to trace amounts of impurities in both the as-grown pristine and Zr-doped ThO$_{2}$ crystals.

In our previous theoretical investigations, we predicted the effects of vacancies, substitutional defects, and Schottky defects using T-matrix Green function method \cite{Malakkal_PRM}. However, direct experimental comparison of individual fission products on the thermal conductivity of ThO$_{2}$ was lacking. In this study, we have, for the first time, experimentally measured the thermal conductivity of 1 at$\%$ Zr-doped ThO$_{2}$ and compared it with theoretical predictions by T-matrix over the temperature range of 77-300 K. We also accounted for isotopic scattering in addition to phonon-defect scattering. The predictions from 100-300 K exhibit excellent agreement with the experimental measurements, strongly suggesting that the theoretical framework employed in this work effectively predicts the influence of fission products on the thermal conductivity of nuclear fuel. However, despite adding isotopic-phonon scattering corrections, the predicted results below 100 K for Zr-doped ThO$_{2}$ were slightly overestimated as compared to the experimental measurement. Notably, when only the perturbation due to mass-mismatch was considered for predicting thermal conductivity, the reduction in thermal conductivity was underpredicted. This indicates the need for a more advanced T-matrix approach.

The reason for the discrepancy at temperatures below 100 K is unclear. Potential causes in the theoretical calculations include inadequate mesh size to accurately capture low-frequency phonon populations that dominate at low temperatures and unaccounted native defects. One of the unaccounted defects that might be present in the Zr-doped ThO$_{2}$  sample is Hf impurity, which is usually present in Zr materials. Additionally, the assumption that all phonon spectra are excited might not hold true for real samples, where only certain phonon modes activate at low temperatures, leading to higher predicted thermal conductivities than experimental measurements. Despite the slight overprediction below 100 K, our study suggests that the nonperturbative Green's function methodology to compute the phonon-point defect scattering rates that consider the local distortion around the point defect—including the mass difference changes, interatomic force constants, and structural relaxation near the point defects—is an excellent method for predicting the role of fission products in actinide materials' thermal conductivity.

To perform a more quantitative comparison between experiment and a first principles model, we use an approach based on classical analytical expressions for thermal conductivity by Klemens \cite{hurley2022thermal,Klemens_1955}. This approach is typically used as a baseline for thermal conductivity correlations in fuel performance modeling \cite{RONCHI_2004,kato2024}. Recently, this has become a practical approach to connect results of atomistic modeling and to fuel performance analysis \cite{JIN2022Impact,FERRIGNO_2023}. Thermal conductivity correlations  define a phonon scattering parameter for each substitutional fission product or impurity $\Gamma_i=C_iS_i^2$, where $C_i$ is impurity concentration. The scattering cross-section is defined in terms of ionic mismatch using \cite{FUKUSHIMA1986,DURIEZ2000}:
\begin{equation}
    S^2=\left(\frac{\Delta m}{m}\right)^2 + 
    \epsilon\left(\frac{\Delta r}{r}\right)^2
\end{equation}
where ionic mass $\Delta m$ and radius $\Delta r$ are mismatch between impurity and host atoms. $epsilon$ is mainly used as fitting parameter. Essentially, while this approach is inspired by established classical thermal transport model, but still relies on empirical parameterization based on experimental results.

We fit our experimental results using conductivity model outlined in \cite{chauhan2021} and ThO$_2$ parameters from \cite{hua2023thermal} to determine $S^2=0.21$. Assuming Zr substitutes Th, using the $m_{Th}=232$ and $m_{Zr}=91.2$ and as ionic masses and radia $r_{Th}=0.105$ and $r_{Zr}=0.084$ \cite{Shannon:a12967}, and a typical value for $epsilon=100$, we find a significantly  larger value $S^2=4.36$ than experiment. A more plausible result is obtained if use an ionic radia determined by Horii et al. for MOX \cite{HORII_2024_JNM}, which estimates $S^2=1.09$.  On the other hand, fitting of our first-principles data results in value $S^2=0.4$ that provides a much a better value compared to our experimental results. It is important to note that the KCM model fails to capture the low-temperature conductivity. 
\color{black}

The successful prediction of thermal conductivity degradation due to Zr substitution in ThO$_{2}$ using first principles techniques, followed by experimental confirmation, may pave the way for the application of the first principles approach to accurately predict the impact of point defects created by other fission products generated in the microstructure of thorium dioxide on its thermal conductivity at operational temperatures. This approach has the potential to set the stage for the incorporation of high-fidelity thermal conductivity degradation models into fuel performance codes such as BISON \cite{Bison}. Furthermore, this work can lay the foundation for extending these methodologies to other next-generation nuclear fuel materials. The fundamental understanding of thermal conductivity degradation gained from this study can be instrumental in refining and enhancing the predictive capabilities of thermal conductivity degradation models. Such improvements in fuel performance codes are crucial for optimizing the safety, efficiency, and longevity of nuclear reactors.

\section{Conclusion}
In conclusion, we have fabricated Zr-doped ThO$_{2}$ using hydrothermal methods to simulate the effect of substitutional defects analogous to those that occur during neutron irradiation in the nuclear fuel fission process. The fabricated single crystals were characterized by XRF, Raman spectroscopy, and PL measurements to verify the quality and integrity of the sample. Raman spectroscopy revealed that the substitutional doping of ThO$_{2}$ with 1 at\%  Zr did not significantly alter the Raman active  T$_2$g mode. The PL spectra also did not reveal notable differences between the pristine and Zr-doped ThO$_{2}$. Thermal conductivity measurements were conducted using a spatial domain thermo-reflectance technique, and the results were compared with the predicted thermal conductivity values using the Green function T-matrix method. The predicted decrease in thermal conductivity due to Zr doping was in excellent agreement with measured values. These findings emphasize the need for computational models incorporating non-perturbative methods to accurately predict the effects of substitutional defects on the thermal conductivity of ThO$_{2}$. These insights contribute to understanding of defect-related thermal conductivity degradation in advanced nuclear fuels and the development of more accurate fuel performance codes.

\section{ACKNOWLEDGMENTS}
This work was supported by the Center for Thermal Energy Transport under Irradiation (TETI), an Energy Frontier Research Center (EFRC) funded by the U.S. Department of Energy, Office of Science, and Office of Basic Energy Sciences. The authors also acknowledge that this research made use of the resources of the High Performance Computing Center at Idaho National Laboratory, which is supported by the Office of Nuclear Energy of the U.S. Department of Energy and the Nuclear Science User Facilities under Contract No. DE-AC07-05ID14517. This manuscript has been authored by Battelle Energy Alliance, LLC under Contract No. DE-AC07-05ID14517 with the U.S. Department of Energy. The United States Government retains and the publisher, by accepting the article for publication, acknowledges that the U.S. Government retains a nonexclusive, paid-up, irrevocable, world-wide license to publish or reproduce the published form of this manuscript, or allow others to do so, for U.S. Government purposes.

\bibliography{ref}

\providecommand{\noopsort}[1]{}\providecommand{\singleletter}[1]{#1}%
\begin{thebibliography}{72}%
\makeatletter
\providecommand \@ifxundefined [1]{%
 \@ifx{#1\undefined}
}%
\providecommand \@ifnum [1]{%
 \ifnum #1\expandafter \@firstoftwo
 \else \expandafter \@secondoftwo
 \fi
}%
\providecommand \@ifx [1]{%
 \ifx #1\expandafter \@firstoftwo
 \else \expandafter \@secondoftwo
 \fi
}%
\providecommand \natexlab [1]{#1}%
\providecommand \enquote  [1]{``#1''}%
\providecommand \bibnamefont  [1]{#1}%
\providecommand \bibfnamefont [1]{#1}%
\providecommand \citenamefont [1]{#1}%
\providecommand \href@noop [0]{\@secondoftwo}%
\providecommand \href [0]{\begingroup \@sanitize@url \@href}%
\providecommand \@href[1]{\@@startlink{#1}\@@href}%
\providecommand \@@href[1]{\endgroup#1\@@endlink}%
\providecommand \@sanitize@url [0]{\catcode `\\12\catcode `\$12\catcode
  `\&12\catcode `\#12\catcode `\^12\catcode `\_12\catcode `\%12\relax}%
\providecommand \@@startlink[1]{}%
\providecommand \@@endlink[0]{}%
\providecommand \url  [0]{\begingroup\@sanitize@url \@url }%
\providecommand \@url [1]{\endgroup\@href {#1}{\urlprefix }}%
\providecommand \urlprefix  [0]{URL }%
\providecommand \Eprint [0]{\href }%
\providecommand \doibase [0]{https://doi.org/}%
\providecommand \selectlanguage [0]{\@gobble}%
\providecommand \bibinfo  [0]{\@secondoftwo}%
\providecommand \bibfield  [0]{\@secondoftwo}%
\providecommand \translation [1]{[#1]}%
\providecommand \BibitemOpen [0]{}%
\providecommand \bibitemStop [0]{}%
\providecommand \bibitemNoStop [0]{.\EOS\space}%
\providecommand \EOS [0]{\spacefactor3000\relax}%
\providecommand \BibitemShut  [1]{\csname bibitem#1\endcsname}%
\let\auto@bib@innerbib\@empty
\bibitem [{\citenamefont {Herring}\ \emph {et~al.}(2001)\citenamefont
  {Herring}, \citenamefont {MacDonald}, \citenamefont {Weaver},\ and\
  \citenamefont {Kullberg}}]{Herring2001}%
  \BibitemOpen
  \bibfield  {author} {\bibinfo {author} {\bibfnamefont {J.~S.}\ \bibnamefont
  {Herring}}, \bibinfo {author} {\bibfnamefont {P.~E.}\ \bibnamefont
  {MacDonald}}, \bibinfo {author} {\bibfnamefont {K.~D.}\ \bibnamefont
  {Weaver}},\ and\ \bibinfo {author} {\bibfnamefont {C.}~\bibnamefont
  {Kullberg}},\ }\bibfield  {title} {\bibinfo {title} {Low cost, proliferation
  resistant, uranium-thorium dioxide fuels for light water reactors},\
  }\href@noop {} {\bibfield  {journal} {\bibinfo  {journal} {Nuclear
  Engineering and Design}\ }\textbf {\bibinfo {volume} {203}},\ \bibinfo
  {pages} {65 } (\bibinfo {year} {2001})}\BibitemShut {NoStop}%
\bibitem [{\citenamefont {Hania}\ and\ \citenamefont
  {Klaassen}(2012)}]{Hania2012304T}%
  \BibitemOpen
  \bibfield  {author} {\bibinfo {author} {\bibfnamefont {P.~R.}\ \bibnamefont
  {Hania}}\ and\ \bibinfo {author} {\bibfnamefont {F.~C.}\ \bibnamefont
  {Klaassen}},\ }\bibfield  {title} {\bibinfo {title} {3.04 – thorium oxide
  fuel}\ }(\bibinfo {year} {2012})\BibitemShut {NoStop}%
\bibitem [{\citenamefont {Ashley}\ \emph {et~al.}(2012)\citenamefont {Ashley},
  \citenamefont {Parks}, \citenamefont {Nuttall}, \citenamefont {Boxall},\ and\
  \citenamefont {Grimes}}]{Ashley_2012}%
  \BibitemOpen
  \bibfield  {author} {\bibinfo {author} {\bibfnamefont {S.~F.}\ \bibnamefont
  {Ashley}}, \bibinfo {author} {\bibfnamefont {G.~T.}\ \bibnamefont {Parks}},
  \bibinfo {author} {\bibfnamefont {W.~J.}\ \bibnamefont {Nuttall}}, \bibinfo
  {author} {\bibfnamefont {C.}~\bibnamefont {Boxall}},\ and\ \bibinfo {author}
  {\bibfnamefont {R.~W.}\ \bibnamefont {Grimes}},\ }\bibfield  {title}
  {\bibinfo {title} {Thorium fuel has risks},\ }\href@noop {} {\bibfield
  {journal} {\bibinfo  {journal} {Nature}\ }\textbf {\bibinfo {volume} {492}},\
  \bibinfo {pages} {31} (\bibinfo {year} {2012})}\BibitemShut {NoStop}%
\bibitem [{\citenamefont {Kutty}\ \emph {et~al.}(2008)\citenamefont {Kutty},
  \citenamefont {Nair}, \citenamefont {Sengupta}, \citenamefont {Basak},
  \citenamefont {Kumar},\ and\ \citenamefont {Kamath}}]{KUTTY_2008}%
  \BibitemOpen
  \bibfield  {author} {\bibinfo {author} {\bibfnamefont {T.}~\bibnamefont
  {Kutty}}, \bibinfo {author} {\bibfnamefont {M.}~\bibnamefont {Nair}},
  \bibinfo {author} {\bibfnamefont {P.}~\bibnamefont {Sengupta}}, \bibinfo
  {author} {\bibfnamefont {U.}~\bibnamefont {Basak}}, \bibinfo {author}
  {\bibfnamefont {A.}~\bibnamefont {Kumar}},\ and\ \bibinfo {author}
  {\bibfnamefont {H.}~\bibnamefont {Kamath}},\ }\bibfield  {title} {\bibinfo
  {title} {Characterization of ({T}h,{U}){O}$_2$ fuel pellets made by
  impregnation technique},\ }\href@noop {} {\bibfield  {journal} {\bibinfo
  {journal} {Journal of Nuclear Materials}\ }\textbf {\bibinfo {volume}
  {374}},\ \bibinfo {pages} {9} (\bibinfo {year} {2008})}\BibitemShut {NoStop}%
\bibitem [{\citenamefont {Hurley}\ \emph {et~al.}(2022)\citenamefont {Hurley},
  \citenamefont {El-Azab}, \citenamefont {Bryan}, \citenamefont {Cooper},
  \citenamefont {Dennett}, \citenamefont {Gofryk}, \citenamefont {He},
  \citenamefont {Khafizov}, \citenamefont {Lander}, \citenamefont {Manley},
  \citenamefont {Mann}, \citenamefont {Marianetti}, \citenamefont {Rickert},
  \citenamefont {Selim}, \citenamefont {Tonks},\ and\ \citenamefont
  {Wharry}}]{hurley2022thermal}%
  \BibitemOpen
  \bibfield  {author} {\bibinfo {author} {\bibfnamefont {D.~H.}\ \bibnamefont
  {Hurley}}, \bibinfo {author} {\bibfnamefont {A.}~\bibnamefont {El-Azab}},
  \bibinfo {author} {\bibfnamefont {M.~S.}\ \bibnamefont {Bryan}}, \bibinfo
  {author} {\bibfnamefont {M.~W.~D.}\ \bibnamefont {Cooper}}, \bibinfo {author}
  {\bibfnamefont {C.~A.}\ \bibnamefont {Dennett}}, \bibinfo {author}
  {\bibfnamefont {K.}~\bibnamefont {Gofryk}}, \bibinfo {author} {\bibfnamefont
  {L.}~\bibnamefont {He}}, \bibinfo {author} {\bibfnamefont {M.}~\bibnamefont
  {Khafizov}}, \bibinfo {author} {\bibfnamefont {G.~H.}\ \bibnamefont
  {Lander}}, \bibinfo {author} {\bibfnamefont {M.~E.}\ \bibnamefont {Manley}},
  \bibinfo {author} {\bibfnamefont {J.~M.}\ \bibnamefont {Mann}}, \bibinfo
  {author} {\bibfnamefont {C.~A.}\ \bibnamefont {Marianetti}}, \bibinfo
  {author} {\bibfnamefont {K.}~\bibnamefont {Rickert}}, \bibinfo {author}
  {\bibfnamefont {F.~A.}\ \bibnamefont {Selim}}, \bibinfo {author}
  {\bibfnamefont {M.~R.}\ \bibnamefont {Tonks}},\ and\ \bibinfo {author}
  {\bibfnamefont {J.~P.}\ \bibnamefont {Wharry}},\ }\bibfield  {title}
  {\bibinfo {title} {Thermal energy transport in oxide nuclear fuel},\
  }\href@noop {} {\bibfield  {journal} {\bibinfo  {journal} {Chemical Reviews}\
  }\textbf {\bibinfo {volume} {122}},\ \bibinfo {pages} {3711} (\bibinfo {year}
  {2022})}\BibitemShut {NoStop}%
\bibitem [{iae(2005)}]{iaea2005}%
  \BibitemOpen
  \href
  {https://www.iaea.org/publications/7192/thorium-fuel-cycle-potential-benefits-and-challenges}
  {\emph {\bibinfo {title} {Thorium Fuel Cycle - Potential Benefits and
  Challenges}}},\ \bibinfo {series} {TECDOC Series}\ No.\ \bibinfo {number}
  {1450}\ (\bibinfo  {publisher} {INTERNATIONAL ATOMIC ENERGY AGENCY},\
  \bibinfo {address} {Vienna},\ \bibinfo {year} {2005})\BibitemShut {NoStop}%
\bibitem [{\citenamefont {Deskins}\ \emph {et~al.}(2022)\citenamefont
  {Deskins}, \citenamefont {Khanolkar}, \citenamefont {Mazumder}, \citenamefont
  {Dennett}, \citenamefont {Bawane}, \citenamefont {Hua}, \citenamefont
  {Ferrigno}, \citenamefont {He}, \citenamefont {Mann}, \citenamefont
  {Khafizov}, \citenamefont {Hurley},\ and\ \citenamefont
  {El-Azab}}]{DESKINS2022}%
  \BibitemOpen
  \bibfield  {author} {\bibinfo {author} {\bibfnamefont {W.~R.}\ \bibnamefont
  {Deskins}}, \bibinfo {author} {\bibfnamefont {A.}~\bibnamefont {Khanolkar}},
  \bibinfo {author} {\bibfnamefont {S.}~\bibnamefont {Mazumder}}, \bibinfo
  {author} {\bibfnamefont {C.~A.}\ \bibnamefont {Dennett}}, \bibinfo {author}
  {\bibfnamefont {K.}~\bibnamefont {Bawane}}, \bibinfo {author} {\bibfnamefont
  {Z.}~\bibnamefont {Hua}}, \bibinfo {author} {\bibfnamefont {J.}~\bibnamefont
  {Ferrigno}}, \bibinfo {author} {\bibfnamefont {L.}~\bibnamefont {He}},
  \bibinfo {author} {\bibfnamefont {J.~M.}\ \bibnamefont {Mann}}, \bibinfo
  {author} {\bibfnamefont {M.}~\bibnamefont {Khafizov}}, \bibinfo {author}
  {\bibfnamefont {D.~H.}\ \bibnamefont {Hurley}},\ and\ \bibinfo {author}
  {\bibfnamefont {A.}~\bibnamefont {El-Azab}},\ }\bibfield  {title} {\bibinfo
  {title} {A combined theoretical-experimental investigation of thermal
  transport in low-dose irradiated thorium dioxide},\ }\href
  {https://doi.org/https://doi.org/10.1016/j.actamat.2022.118379} {\bibfield
  {journal} {\bibinfo  {journal} {Acta Materialia}\ }\textbf {\bibinfo {volume}
  {241}},\ \bibinfo {pages} {118379} (\bibinfo {year} {2022})}\BibitemShut
  {NoStop}%
\bibitem [{\citenamefont {Malakkal}\ \emph {et~al.}(2019)\citenamefont
  {Malakkal}, \citenamefont {Prasad}, \citenamefont {Jossou}, \citenamefont
  {Ranasinghe}, \citenamefont {Szpunar}, \citenamefont {Bichler},\ and\
  \citenamefont {Szpunar}}]{MALAKKAL2019507}%
  \BibitemOpen
  \bibfield  {author} {\bibinfo {author} {\bibfnamefont {L.}~\bibnamefont
  {Malakkal}}, \bibinfo {author} {\bibfnamefont {A.}~\bibnamefont {Prasad}},
  \bibinfo {author} {\bibfnamefont {E.}~\bibnamefont {Jossou}}, \bibinfo
  {author} {\bibfnamefont {J.}~\bibnamefont {Ranasinghe}}, \bibinfo {author}
  {\bibfnamefont {B.}~\bibnamefont {Szpunar}}, \bibinfo {author} {\bibfnamefont
  {L.}~\bibnamefont {Bichler}},\ and\ \bibinfo {author} {\bibfnamefont
  {J.}~\bibnamefont {Szpunar}},\ }\bibfield  {title} {\bibinfo {title} {Thermal
  conductivity of bulk and porous \uppercase{T}h\uppercase{O}$_2$: Atomistic
  and experimental study},\ }\href
  {https://doi.org/https://doi.org/10.1016/j.jallcom.2019.05.274} {\bibfield
  {journal} {\bibinfo  {journal} {Journal of Alloys and Compounds}\ }\textbf
  {\bibinfo {volume} {798}},\ \bibinfo {pages} {507} (\bibinfo {year}
  {2019})}\BibitemShut {NoStop}%
\bibitem [{\citenamefont {Worrall}\ \emph {et~al.}(2023)\citenamefont
  {Worrall}, \citenamefont {Woolstenhulme},\ and\ \citenamefont
  {Turner}}]{Clean_energy}%
  \BibitemOpen
  \bibfield  {author} {\bibinfo {author} {\bibfnamefont {M.}~\bibnamefont
  {Worrall}}, \bibinfo {author} {\bibfnamefont {N.}~\bibnamefont
  {Woolstenhulme}},\ and\ \bibinfo {author} {\bibfnamefont {C.}~\bibnamefont
  {Turner}},\ }\href@noop {} {\emph {\bibinfo {title} {Final {CRADA} Report:
  Accelerated Burn-up Accumulation Test of Clean Core Thorium Energy Designated
  {ANEEL} Fuel}}}\ (\bibinfo {year} {2023})\BibitemShut {NoStop}%
\bibitem [{\citenamefont {of~Mines}(2020)}]{IREL_indian2020}%
  \BibitemOpen
  \bibfield  {author} {\bibinfo {author} {\bibfnamefont {I.~B.}\ \bibnamefont
  {of~Mines}},\ }\bibfield  {title} {\bibinfo {title} {Indian minerals yearbook
  2019 (part-iii: Mineral reviews) rare earths},\ }\href@noop {} {\bibfield
  {journal} {\bibinfo  {journal} {Indian Bureau of Mines}\ } (\bibinfo {year}
  {2020})}\BibitemShut {NoStop}%
\bibitem [{\citenamefont {Olander}(1976)}]{Olander_1976}%
  \BibitemOpen
  \bibfield  {author} {\bibinfo {author} {\bibfnamefont {D.~R.}\ \bibnamefont
  {Olander}},\ }\bibfield  {title} {\bibinfo {title} {Fundamental aspects of
  nuclear reactor fuel elements}\ }\href {https://doi.org/10.2172/7343826}
  {10.2172/7343826} (\bibinfo {year} {1976})\BibitemShut {NoStop}%
\bibitem [{\citenamefont {Kawano}\ \emph {et~al.}(2023)\citenamefont {Kawano},
  \citenamefont {Randrup}, \citenamefont {Schunck}, \citenamefont {Talou},\
  and\ \citenamefont {Tovesson}}]{Kawano2023}%
  \BibitemOpen
  \bibfield  {author} {\bibinfo {author} {\bibfnamefont {T.}~\bibnamefont
  {Kawano}}, \bibinfo {author} {\bibfnamefont {J.}~\bibnamefont {Randrup}},
  \bibinfo {author} {\bibfnamefont {N.}~\bibnamefont {Schunck}}, \bibinfo
  {author} {\bibfnamefont {P.}~\bibnamefont {Talou}},\ and\ \bibinfo {author}
  {\bibfnamefont {F.}~\bibnamefont {Tovesson}},\ }\bibinfo {title} {Fission
  fragments and fission products},\ in\ \href
  {https://doi.org/10.1007/978-3-031-14545-2_2} {\emph {\bibinfo {booktitle}
  {Nuclear Fission: Theories, Experiments and Applications}}},\ \bibinfo
  {editor} {edited by\ \bibinfo {editor} {\bibfnamefont {P.}~\bibnamefont
  {Talou}}\ and\ \bibinfo {editor} {\bibfnamefont {R.}~\bibnamefont {Vogt}}}\
  (\bibinfo  {publisher} {Springer International Publishing},\ \bibinfo
  {address} {Cham},\ \bibinfo {year} {2023})\ pp.\ \bibinfo {pages}
  {141--262}\BibitemShut {NoStop}%
\bibitem [{\citenamefont {Williamson}\ \emph {et~al.}(2021)\citenamefont
  {Williamson}, \citenamefont {Hales}, \citenamefont {Novascone}, \citenamefont
  {Pastore}, \citenamefont {Gamble}, \citenamefont {Spencer}, \citenamefont
  {Jiang}, \citenamefont {Pitts}, \citenamefont {Casagranda}, \citenamefont
  {Schwen}, \citenamefont {Zabriskie}, \citenamefont {Toptan}, \citenamefont
  {Gardner}, \citenamefont {Matthews}, \citenamefont {Liu},\ and\ \citenamefont
  {Chen}}]{Bison}%
  \BibitemOpen
  \bibfield  {author} {\bibinfo {author} {\bibfnamefont {R.~L.}\ \bibnamefont
  {Williamson}}, \bibinfo {author} {\bibfnamefont {J.~D.}\ \bibnamefont
  {Hales}}, \bibinfo {author} {\bibfnamefont {S.~R.}\ \bibnamefont
  {Novascone}}, \bibinfo {author} {\bibfnamefont {G.}~\bibnamefont {Pastore}},
  \bibinfo {author} {\bibfnamefont {K.~A.}\ \bibnamefont {Gamble}}, \bibinfo
  {author} {\bibfnamefont {B.~W.}\ \bibnamefont {Spencer}}, \bibinfo {author}
  {\bibfnamefont {W.}~\bibnamefont {Jiang}}, \bibinfo {author} {\bibfnamefont
  {S.~A.}\ \bibnamefont {Pitts}}, \bibinfo {author} {\bibfnamefont
  {A.}~\bibnamefont {Casagranda}}, \bibinfo {author} {\bibfnamefont
  {D.}~\bibnamefont {Schwen}}, \bibinfo {author} {\bibfnamefont {A.~X.}\
  \bibnamefont {Zabriskie}}, \bibinfo {author} {\bibfnamefont {A.}~\bibnamefont
  {Toptan}}, \bibinfo {author} {\bibfnamefont {R.}~\bibnamefont {Gardner}},
  \bibinfo {author} {\bibfnamefont {C.}~\bibnamefont {Matthews}}, \bibinfo
  {author} {\bibfnamefont {W.}~\bibnamefont {Liu}},\ and\ \bibinfo {author}
  {\bibfnamefont {H.}~\bibnamefont {Chen}},\ }\bibfield  {title} {\bibinfo
  {title} {Bison: A flexible code for advanced simulation of the performance of
  multiple nuclear fuel forms},\ }\href
  {https://doi.org/10.1080/00295450.2020.1836940} {\bibfield  {journal}
  {\bibinfo  {journal} {Nuclear Technology}\ }\textbf {\bibinfo {volume}
  {207}},\ \bibinfo {pages} {954} (\bibinfo {year} {2021})}\BibitemShut
  {NoStop}%
\bibitem [{\citenamefont {Magni}\ \emph {et~al.}(2021)\citenamefont {Magni},
  \citenamefont {{Del Nevo}}, \citenamefont {Luzzi}, \citenamefont {Rozzia},
  \citenamefont {Adorni}, \citenamefont {Schubert},\ and\ \citenamefont {{Van
  Uffelen}}}]{Transuranus_MAGNI_2021}%
  \BibitemOpen
  \bibfield  {author} {\bibinfo {author} {\bibfnamefont {A.}~\bibnamefont
  {Magni}}, \bibinfo {author} {\bibfnamefont {A.}~\bibnamefont {{Del Nevo}}},
  \bibinfo {author} {\bibfnamefont {L.}~\bibnamefont {Luzzi}}, \bibinfo
  {author} {\bibfnamefont {D.}~\bibnamefont {Rozzia}}, \bibinfo {author}
  {\bibfnamefont {M.}~\bibnamefont {Adorni}}, \bibinfo {author} {\bibfnamefont
  {A.}~\bibnamefont {Schubert}},\ and\ \bibinfo {author} {\bibfnamefont
  {P.}~\bibnamefont {{Van Uffelen}}},\ }\bibfield  {title} {\bibinfo {title}
  {Chapter 8 - the transuranus fuel performance code},\ }in\ \href
  {https://doi.org/https://doi.org/10.1016/B978-0-12-818190-4.00008-5} {\emph
  {\bibinfo {booktitle} {Nuclear Power Plant Design and Analysis Codes}}},\
  \bibinfo {series and number} {Woodhead Publishing Series in Energy},\
  \bibinfo {editor} {edited by\ \bibinfo {editor} {\bibfnamefont
  {J.}~\bibnamefont {Wang}}, \bibinfo {editor} {\bibfnamefont {X.}~\bibnamefont
  {Li}}, \bibinfo {editor} {\bibfnamefont {C.}~\bibnamefont {Allison}},\ and\
  \bibinfo {editor} {\bibfnamefont {J.}~\bibnamefont {Hohorst}}}\ (\bibinfo
  {publisher} {Woodhead Publishing},\ \bibinfo {year} {2021})\ pp.\ \bibinfo
  {pages} {161--205}\BibitemShut {NoStop}%
\bibitem [{\citenamefont {Introïni}\ \emph {et~al.}(2024)\citenamefont
  {Introïni}, \citenamefont {Ramière}, \citenamefont {Sercombe},
  \citenamefont {Michel}, \citenamefont {Helfer},\ and\ \citenamefont
  {Fauque}}]{ALCYONE_INTROINI_2024}%
  \BibitemOpen
  \bibfield  {author} {\bibinfo {author} {\bibfnamefont {C.}~\bibnamefont
  {Introïni}}, \bibinfo {author} {\bibfnamefont {I.}~\bibnamefont {Ramière}},
  \bibinfo {author} {\bibfnamefont {J.}~\bibnamefont {Sercombe}}, \bibinfo
  {author} {\bibfnamefont {B.}~\bibnamefont {Michel}}, \bibinfo {author}
  {\bibfnamefont {T.}~\bibnamefont {Helfer}},\ and\ \bibinfo {author}
  {\bibfnamefont {J.}~\bibnamefont {Fauque}},\ }\bibfield  {title} {\bibinfo
  {title} {Alcyone: the fuel performance code of the pleiades platform
  dedicated to pwr fuel rods behavior},\ }\href
  {https://doi.org/https://doi.org/10.1016/j.anucene.2024.110711} {\bibfield
  {journal} {\bibinfo  {journal} {Annals of Nuclear Energy}\ }\textbf {\bibinfo
  {volume} {207}},\ \bibinfo {pages} {110711} (\bibinfo {year}
  {2024})}\BibitemShut {NoStop}%
\bibitem [{\citenamefont {{Van Uffelen}}\ \emph {et~al.}(2019)\citenamefont
  {{Van Uffelen}}, \citenamefont {Hales}, \citenamefont {Li}, \citenamefont
  {Rossiter},\ and\ \citenamefont {Williamson}}]{VANUFFELEN_2019_review}%
  \BibitemOpen
  \bibfield  {author} {\bibinfo {author} {\bibfnamefont {P.}~\bibnamefont {{Van
  Uffelen}}}, \bibinfo {author} {\bibfnamefont {J.}~\bibnamefont {Hales}},
  \bibinfo {author} {\bibfnamefont {W.}~\bibnamefont {Li}}, \bibinfo {author}
  {\bibfnamefont {G.}~\bibnamefont {Rossiter}},\ and\ \bibinfo {author}
  {\bibfnamefont {R.}~\bibnamefont {Williamson}},\ }\bibfield  {title}
  {\bibinfo {title} {A review of fuel performance modelling},\ }\href
  {https://doi.org/https://doi.org/10.1016/j.jnucmat.2018.12.037} {\bibfield
  {journal} {\bibinfo  {journal} {Journal of Nuclear Materials}\ }\textbf
  {\bibinfo {volume} {516}},\ \bibinfo {pages} {373} (\bibinfo {year}
  {2019})}\BibitemShut {NoStop}%
\bibitem [{\citenamefont {Lucuta}\ \emph {et~al.}(1996)\citenamefont {Lucuta},
  \citenamefont {Matzke},\ and\ \citenamefont {Hastings}}]{LUCUTA_1996}%
  \BibitemOpen
  \bibfield  {author} {\bibinfo {author} {\bibfnamefont {P.}~\bibnamefont
  {Lucuta}}, \bibinfo {author} {\bibfnamefont {H.}~\bibnamefont {Matzke}},\
  and\ \bibinfo {author} {\bibfnamefont {I.}~\bibnamefont {Hastings}},\
  }\bibfield  {title} {\bibinfo {title} {A pragmatic approach to modelling
  thermal conductivity of irradiated {UO}$_2$ fuel: Review and
  recommendations},\ }\href
  {https://doi.org/https://doi.org/10.1016/S0022-3115(96)00404-7} {\bibfield
  {journal} {\bibinfo  {journal} {Journal of Nuclear Materials}\ }\textbf
  {\bibinfo {volume} {232}},\ \bibinfo {pages} {166} (\bibinfo {year}
  {1996})}\BibitemShut {NoStop}%
\bibitem [{\citenamefont {Ferrigno}\ \emph {et~al.}(2024)\citenamefont
  {Ferrigno}, \citenamefont {Pavlov}, \citenamefont {Poudel}, \citenamefont
  {Salvato}, \citenamefont {Tsai}, \citenamefont {Merritt}, \citenamefont
  {Hansen}, \citenamefont {Munro}, \citenamefont {Cappia},\ and\ \citenamefont
  {Khafizov}}]{FERRIGNO_2024}%
  \BibitemOpen
  \bibfield  {author} {\bibinfo {author} {\bibfnamefont {J.}~\bibnamefont
  {Ferrigno}}, \bibinfo {author} {\bibfnamefont {T.}~\bibnamefont {Pavlov}},
  \bibinfo {author} {\bibfnamefont {N.}~\bibnamefont {Poudel}}, \bibinfo
  {author} {\bibfnamefont {D.}~\bibnamefont {Salvato}}, \bibinfo {author}
  {\bibfnamefont {C.}~\bibnamefont {Tsai}}, \bibinfo {author} {\bibfnamefont
  {B.}~\bibnamefont {Merritt}}, \bibinfo {author} {\bibfnamefont
  {A.}~\bibnamefont {Hansen}}, \bibinfo {author} {\bibfnamefont
  {T.}~\bibnamefont {Munro}}, \bibinfo {author} {\bibfnamefont
  {F.}~\bibnamefont {Cappia}},\ and\ \bibinfo {author} {\bibfnamefont
  {M.}~\bibnamefont {Khafizov}},\ }\bibfield  {title} {\bibinfo {title}
  {Analysis of radially resolved thermal conductivity in high burnup mixed
  oxide fuel and comparison to thermal conductivity correlations implemented in
  fuel performance codes},\ }\href
  {https://doi.org/https://doi.org/10.1016/j.jnucmat.2024.155090} {\bibfield
  {journal} {\bibinfo  {journal} {Journal of Nuclear Materials}\ }\textbf
  {\bibinfo {volume} {596}},\ \bibinfo {pages} {155090} (\bibinfo {year}
  {2024})}\BibitemShut {NoStop}%
\bibitem [{\citenamefont {Magni}\ \emph {et~al.}(2020)\citenamefont {Magni},
  \citenamefont {Barani}, \citenamefont {{Del Nevo}}, \citenamefont {Pizzocri},
  \citenamefont {Staicu}, \citenamefont {{Van Uffelen}},\ and\ \citenamefont
  {Luzzi}}]{MAGNI2020152410}%
  \BibitemOpen
  \bibfield  {author} {\bibinfo {author} {\bibfnamefont {A.}~\bibnamefont
  {Magni}}, \bibinfo {author} {\bibfnamefont {T.}~\bibnamefont {Barani}},
  \bibinfo {author} {\bibfnamefont {A.}~\bibnamefont {{Del Nevo}}}, \bibinfo
  {author} {\bibfnamefont {D.}~\bibnamefont {Pizzocri}}, \bibinfo {author}
  {\bibfnamefont {D.}~\bibnamefont {Staicu}}, \bibinfo {author} {\bibfnamefont
  {P.}~\bibnamefont {{Van Uffelen}}},\ and\ \bibinfo {author} {\bibfnamefont
  {L.}~\bibnamefont {Luzzi}},\ }\bibfield  {title} {\bibinfo {title} {Modelling
  and assessment of thermal conductivity and melting behaviour of mox fuel for
  fast reactor applications},\ }\href
  {https://doi.org/https://doi.org/10.1016/j.jnucmat.2020.152410} {\bibfield
  {journal} {\bibinfo  {journal} {Journal of Nuclear Materials}\ }\textbf
  {\bibinfo {volume} {541}},\ \bibinfo {pages} {152410} (\bibinfo {year}
  {2020})}\BibitemShut {NoStop}%
\bibitem [{\citenamefont {Ferrigno}\ \emph {et~al.}(2023)\citenamefont
  {Ferrigno}, \citenamefont {Adnan},\ and\ \citenamefont
  {Khafizov}}]{FERRIGNO_2023}%
  \BibitemOpen
  \bibfield  {author} {\bibinfo {author} {\bibfnamefont {J.}~\bibnamefont
  {Ferrigno}}, \bibinfo {author} {\bibfnamefont {S.}~\bibnamefont {Adnan}},\
  and\ \bibinfo {author} {\bibfnamefont {M.}~\bibnamefont {Khafizov}},\
  }\bibfield  {title} {\bibinfo {title} {Influence of point defect accumulation
  on in-pile thermal conductivity degradation: Fuel rod defect distribution and
  deviation between in-pile and post irradiation thermal conductivity},\ }\href
  {https://doi.org/https://doi.org/10.1016/j.jnucmat.2022.154108} {\bibfield
  {journal} {\bibinfo  {journal} {Journal of Nuclear Materials}\ }\textbf
  {\bibinfo {volume} {573}},\ \bibinfo {pages} {154108} (\bibinfo {year}
  {2023})}\BibitemShut {NoStop}%
\bibitem [{\citenamefont {Dennett}\ \emph {et~al.}(2021)\citenamefont
  {Dennett}, \citenamefont {Deskins}, \citenamefont {Khafizov}, \citenamefont
  {Hua}, \citenamefont {Khanolkar}, \citenamefont {Bawane}, \citenamefont {Fu},
  \citenamefont {Mann}, \citenamefont {Marianetti}, \citenamefont {He},
  \citenamefont {Hurley},\ and\ \citenamefont {El-Azab}}]{DENNETT2021}%
  \BibitemOpen
  \bibfield  {author} {\bibinfo {author} {\bibfnamefont {C.~A.}\ \bibnamefont
  {Dennett}}, \bibinfo {author} {\bibfnamefont {W.~R.}\ \bibnamefont
  {Deskins}}, \bibinfo {author} {\bibfnamefont {M.}~\bibnamefont {Khafizov}},
  \bibinfo {author} {\bibfnamefont {Z.}~\bibnamefont {Hua}}, \bibinfo {author}
  {\bibfnamefont {A.}~\bibnamefont {Khanolkar}}, \bibinfo {author}
  {\bibfnamefont {K.}~\bibnamefont {Bawane}}, \bibinfo {author} {\bibfnamefont
  {L.}~\bibnamefont {Fu}}, \bibinfo {author} {\bibfnamefont {J.~M.}\
  \bibnamefont {Mann}}, \bibinfo {author} {\bibfnamefont {C.~A.}\ \bibnamefont
  {Marianetti}}, \bibinfo {author} {\bibfnamefont {L.}~\bibnamefont {He}},
  \bibinfo {author} {\bibfnamefont {D.~H.}\ \bibnamefont {Hurley}},\ and\
  \bibinfo {author} {\bibfnamefont {A.}~\bibnamefont {El-Azab}},\ }\bibfield
  {title} {\bibinfo {title} {An integrated experimental and computational
  investigation of defect and microstructural effects on thermal transport in
  thorium dioxide},\ }\href
  {https://doi.org/https://doi.org/10.1016/j.actamat.2021.116934} {\bibfield
  {journal} {\bibinfo  {journal} {Acta Materialia}\ }\textbf {\bibinfo {volume}
  {213}},\ \bibinfo {pages} {116934} (\bibinfo {year} {2021})}\BibitemShut
  {NoStop}%
\bibitem [{\citenamefont {Liu}\ \emph {et~al.}(2016)\citenamefont {Liu},
  \citenamefont {Cooper}, \citenamefont {McClellan}, \citenamefont {Lashley},
  \citenamefont {Byler}, \citenamefont {Bell}, \citenamefont {Grimes},
  \citenamefont {Stanek},\ and\ \citenamefont {Andersson}}]{Liu_2016_MD}%
  \BibitemOpen
  \bibfield  {author} {\bibinfo {author} {\bibfnamefont {X.-Y.}\ \bibnamefont
  {Liu}}, \bibinfo {author} {\bibfnamefont {M.~W.~D.}\ \bibnamefont {Cooper}},
  \bibinfo {author} {\bibfnamefont {K.~J.}\ \bibnamefont {McClellan}}, \bibinfo
  {author} {\bibfnamefont {J.~C.}\ \bibnamefont {Lashley}}, \bibinfo {author}
  {\bibfnamefont {D.~D.}\ \bibnamefont {Byler}}, \bibinfo {author}
  {\bibfnamefont {B.~D.~C.}\ \bibnamefont {Bell}}, \bibinfo {author}
  {\bibfnamefont {R.~W.}\ \bibnamefont {Grimes}}, \bibinfo {author}
  {\bibfnamefont {C.~R.}\ \bibnamefont {Stanek}},\ and\ \bibinfo {author}
  {\bibfnamefont {D.~A.}\ \bibnamefont {Andersson}},\ }\bibfield  {title}
  {\bibinfo {title} {Molecular dynamics simulation of thermal transport in
  {U}{O}$_2$ containing uranium, oxygen, and fission-product defects},\ }\href
  {https://doi.org/10.1103/PhysRevApplied.6.044015} {\bibfield  {journal}
  {\bibinfo  {journal} {Phys. Rev. Appl.}\ }\textbf {\bibinfo {volume} {6}},\
  \bibinfo {pages} {044015} (\bibinfo {year} {2016})}\BibitemShut {NoStop}%
\bibitem [{\citenamefont {Jin}\ \emph {et~al.}(2022{\natexlab{a}})\citenamefont
  {Jin}, \citenamefont {Dennett}, \citenamefont {Hurley},\ and\ \citenamefont
  {Khafizov}}]{JIN_2022_MD_tho}%
  \BibitemOpen
  \bibfield  {author} {\bibinfo {author} {\bibfnamefont {M.}~\bibnamefont
  {Jin}}, \bibinfo {author} {\bibfnamefont {C.~A.}\ \bibnamefont {Dennett}},
  \bibinfo {author} {\bibfnamefont {D.~H.}\ \bibnamefont {Hurley}},\ and\
  \bibinfo {author} {\bibfnamefont {M.}~\bibnamefont {Khafizov}},\ }\bibfield
  {title} {\bibinfo {title} {Impact of small defects and dislocation loops on
  phonon scattering and thermal transport in {T}h{O}$_2$},\ }\href
  {https://doi.org/https://doi.org/10.1016/j.jnucmat.2022.153758} {\bibfield
  {journal} {\bibinfo  {journal} {Journal of Nuclear Materials}\ }\textbf
  {\bibinfo {volume} {566}},\ \bibinfo {pages} {153758} (\bibinfo {year}
  {2022}{\natexlab{a}})}\BibitemShut {NoStop}%
\bibitem [{\citenamefont {Klemens}(1955)}]{Klemens_1955}%
  \BibitemOpen
  \bibfield  {author} {\bibinfo {author} {\bibfnamefont {P.~G.}\ \bibnamefont
  {Klemens}},\ }\bibfield  {title} {\bibinfo {title} {The scattering of
  low-frequency lattice waves by static imperfections},\ }\href
  {https://doi.org/10.1088/0370-1298/68/12/303} {\bibfield  {journal} {\bibinfo
   {journal} {Proceedings of the Physical Society. Section A}\ }\textbf
  {\bibinfo {volume} {68}},\ \bibinfo {pages} {1113} (\bibinfo {year}
  {1955})}\BibitemShut {NoStop}%
\bibitem [{\citenamefont {Bonev}\ \emph {et~al.}(2023)\citenamefont {Bonev},
  \citenamefont {Chauvin}, \citenamefont {Staicu}, \citenamefont {Dahms},
  \citenamefont {Montagnier}, \citenamefont {Papaioannou}, \citenamefont
  {Dumas}, \citenamefont {Boukhris}, \citenamefont {Viallard}, \citenamefont
  {Lainet}, \citenamefont {Lamontagne},\ and\ \citenamefont
  {Hanifi}}]{BONEV_2023_JNM}%
  \BibitemOpen
  \bibfield  {author} {\bibinfo {author} {\bibfnamefont {P.}~\bibnamefont
  {Bonev}}, \bibinfo {author} {\bibfnamefont {N.}~\bibnamefont {Chauvin}},
  \bibinfo {author} {\bibfnamefont {D.}~\bibnamefont {Staicu}}, \bibinfo
  {author} {\bibfnamefont {E.}~\bibnamefont {Dahms}}, \bibinfo {author}
  {\bibfnamefont {G.}~\bibnamefont {Montagnier}}, \bibinfo {author}
  {\bibfnamefont {D.}~\bibnamefont {Papaioannou}}, \bibinfo {author}
  {\bibfnamefont {J.-C.}\ \bibnamefont {Dumas}}, \bibinfo {author}
  {\bibfnamefont {I.}~\bibnamefont {Boukhris}}, \bibinfo {author}
  {\bibfnamefont {I.}~\bibnamefont {Viallard}}, \bibinfo {author}
  {\bibfnamefont {M.}~\bibnamefont {Lainet}}, \bibinfo {author} {\bibfnamefont
  {J.}~\bibnamefont {Lamontagne}},\ and\ \bibinfo {author} {\bibfnamefont
  {K.}~\bibnamefont {Hanifi}},\ }\bibfield  {title} {\bibinfo {title} {New
  recommendation for the thermal conductivity of irradiated ({U}, {P}u){O}$_2$
  fuels under fast reactor conditions. comparison with recent experimental
  data},\ }\href
  {https://doi.org/https://doi.org/10.1016/j.jnucmat.2023.154326} {\bibfield
  {journal} {\bibinfo  {journal} {Journal of Nuclear Materials}\ }\textbf
  {\bibinfo {volume} {577}},\ \bibinfo {pages} {154326} (\bibinfo {year}
  {2023})}\BibitemShut {NoStop}%
\bibitem [{\citenamefont {Horii}\ \emph {et~al.}(2024)\citenamefont {Horii},
  \citenamefont {Hirooka}, \citenamefont {Uno}, \citenamefont {Ogasawara},
  \citenamefont {Tamura}, \citenamefont {Yamada}, \citenamefont {Furusawa},
  \citenamefont {Murakami},\ and\ \citenamefont {Kato}}]{HORII_2024_JNM}%
  \BibitemOpen
  \bibfield  {author} {\bibinfo {author} {\bibfnamefont {Y.}~\bibnamefont
  {Horii}}, \bibinfo {author} {\bibfnamefont {S.}~\bibnamefont {Hirooka}},
  \bibinfo {author} {\bibfnamefont {H.}~\bibnamefont {Uno}}, \bibinfo {author}
  {\bibfnamefont {M.}~\bibnamefont {Ogasawara}}, \bibinfo {author}
  {\bibfnamefont {T.}~\bibnamefont {Tamura}}, \bibinfo {author} {\bibfnamefont
  {T.}~\bibnamefont {Yamada}}, \bibinfo {author} {\bibfnamefont
  {N.}~\bibnamefont {Furusawa}}, \bibinfo {author} {\bibfnamefont
  {T.}~\bibnamefont {Murakami}},\ and\ \bibinfo {author} {\bibfnamefont
  {M.}~\bibnamefont {Kato}},\ }\bibfield  {title} {\bibinfo {title} {Thermal
  conductivity measurement of uranium–plutonium mixed oxide doped with
  {N}d/{S}m as simulated fission products},\ }\href
  {https://doi.org/https://doi.org/10.1016/j.jnucmat.2023.154799} {\bibfield
  {journal} {\bibinfo  {journal} {Journal of Nuclear Materials}\ }\textbf
  {\bibinfo {volume} {588}},\ \bibinfo {pages} {154799} (\bibinfo {year}
  {2024})}\BibitemShut {NoStop}%
\bibitem [{\citenamefont {Chen}\ \emph {et~al.}(2024)\citenamefont {Chen},
  \citenamefont {Malakkal}, \citenamefont {Khafizov}, \citenamefont {Hurley},\
  and\ \citenamefont {Jin}}]{CHEN_2024}%
  \BibitemOpen
  \bibfield  {author} {\bibinfo {author} {\bibfnamefont {B.}~\bibnamefont
  {Chen}}, \bibinfo {author} {\bibfnamefont {L.}~\bibnamefont {Malakkal}},
  \bibinfo {author} {\bibfnamefont {M.}~\bibnamefont {Khafizov}}, \bibinfo
  {author} {\bibfnamefont {D.~H.}\ \bibnamefont {Hurley}},\ and\ \bibinfo
  {author} {\bibfnamefont {M.}~\bibnamefont {Jin}},\ }\bibfield  {title}
  {\bibinfo {title} {Phonon modal analysis of thermal transport in {T}h{O}$_2$
  with point defects using equilibrium molecular dynamics},\ }\href
  {https://doi.org/https://doi.org/10.1016/j.jnucmat.2024.155314} {\bibfield
  {journal} {\bibinfo  {journal} {Journal of Nuclear Materials}\ }\textbf
  {\bibinfo {volume} {601}},\ \bibinfo {pages} {155314} (\bibinfo {year}
  {2024})}\BibitemShut {NoStop}%
\bibitem [{\citenamefont {Khafizov}\ \emph {et~al.}(2017)\citenamefont
  {Khafizov}, \citenamefont {Chauhan}, \citenamefont {Wang}, \citenamefont
  {Riyad}, \citenamefont {Hang},\ and\ \citenamefont
  {Hurley}}]{Khafizov_2016_investigation}%
  \BibitemOpen
  \bibfield  {author} {\bibinfo {author} {\bibfnamefont {M.}~\bibnamefont
  {Khafizov}}, \bibinfo {author} {\bibfnamefont {V.}~\bibnamefont {Chauhan}},
  \bibinfo {author} {\bibfnamefont {Y.}~\bibnamefont {Wang}}, \bibinfo {author}
  {\bibfnamefont {F.}~\bibnamefont {Riyad}}, \bibinfo {author} {\bibfnamefont
  {N.}~\bibnamefont {Hang}},\ and\ \bibinfo {author} {\bibfnamefont
  {D.}~\bibnamefont {Hurley}},\ }\bibfield  {title} {\bibinfo {title}
  {Investigation of thermal transport in composites and ion beam irradiated
  materials for nuclear energy applications},\ }\href
  {https://doi.org/10.1557/jmr.2016.421} {\bibfield  {journal} {\bibinfo
  {journal} {Journal of Materials Research}\ }\textbf {\bibinfo {volume}
  {32}},\ \bibinfo {pages} {204} (\bibinfo {year} {2017})}\BibitemShut
  {NoStop}%
\bibitem [{\citenamefont {Dennett}\ \emph {et~al.}(2020)\citenamefont
  {Dennett}, \citenamefont {Hua}, \citenamefont {Khanolkar}, \citenamefont
  {Yao}, \citenamefont {Morgan}, \citenamefont {Prusnick}, \citenamefont
  {Poudel}, \citenamefont {French}, \citenamefont {Gofryk}, \citenamefont {He},
  \citenamefont {Shao}, \citenamefont {Khafizov}, \citenamefont {Turner},
  \citenamefont {Mann},\ and\ \citenamefont {Hurley}}]{Dennett_2020}%
  \BibitemOpen
  \bibfield  {author} {\bibinfo {author} {\bibfnamefont {C.~A.}\ \bibnamefont
  {Dennett}}, \bibinfo {author} {\bibfnamefont {Z.}~\bibnamefont {Hua}},
  \bibinfo {author} {\bibfnamefont {A.}~\bibnamefont {Khanolkar}}, \bibinfo
  {author} {\bibfnamefont {T.}~\bibnamefont {Yao}}, \bibinfo {author}
  {\bibfnamefont {P.~K.}\ \bibnamefont {Morgan}}, \bibinfo {author}
  {\bibfnamefont {T.~A.}\ \bibnamefont {Prusnick}}, \bibinfo {author}
  {\bibfnamefont {N.}~\bibnamefont {Poudel}}, \bibinfo {author} {\bibfnamefont
  {A.}~\bibnamefont {French}}, \bibinfo {author} {\bibfnamefont
  {K.}~\bibnamefont {Gofryk}}, \bibinfo {author} {\bibfnamefont
  {L.}~\bibnamefont {He}}, \bibinfo {author} {\bibfnamefont {L.}~\bibnamefont
  {Shao}}, \bibinfo {author} {\bibfnamefont {M.}~\bibnamefont {Khafizov}},
  \bibinfo {author} {\bibfnamefont {D.~B.}\ \bibnamefont {Turner}}, \bibinfo
  {author} {\bibfnamefont {J.~M.}\ \bibnamefont {Mann}},\ and\ \bibinfo
  {author} {\bibfnamefont {D.~H.}\ \bibnamefont {Hurley}},\ }\bibfield  {title}
  {\bibinfo {title} {{The influence of lattice defects, recombination, and
  clustering on thermal transport in single crystal thorium dioxide}},\
  }\href@noop {} {\bibfield  {journal} {\bibinfo  {journal} {APL Materials}\
  }\textbf {\bibinfo {volume} {8}},\ \bibinfo {pages} {111103} (\bibinfo {year}
  {2020})}\BibitemShut {NoStop}%
\bibitem [{\citenamefont {Grimes}(1991)}]{Grimes_1991}%
  \BibitemOpen
  \bibfield  {author} {\bibinfo {author} {\bibfnamefont {R.~W.}\ \bibnamefont
  {Grimes}},\ }\bibinfo {title} {Simulating the behaviour of inert gases in
  {UO}$_2$},\ in\ \href {https://doi.org/10.1007/978-1-4899-3680-6_36} {\emph
  {\bibinfo {booktitle} {Fundamental Aspects of Inert Gases in Solids}}},\
  \bibinfo {editor} {edited by\ \bibinfo {editor} {\bibfnamefont {S.~E.}\
  \bibnamefont {Donnelly}}\ and\ \bibinfo {editor} {\bibfnamefont {J.~H.}\
  \bibnamefont {Evans}}}\ (\bibinfo  {publisher} {Springer US},\ \bibinfo
  {address} {Boston, MA},\ \bibinfo {year} {1991})\ pp.\ \bibinfo {pages}
  {415--429}\BibitemShut {NoStop}%
\bibitem [{\citenamefont {Ronchi}\ \emph {et~al.}(2004)\citenamefont {Ronchi},
  \citenamefont {Sheindlin}, \citenamefont {Staicu},\ and\ \citenamefont
  {Kinoshita}}]{RONCHI_2004}%
  \BibitemOpen
  \bibfield  {author} {\bibinfo {author} {\bibfnamefont {C.}~\bibnamefont
  {Ronchi}}, \bibinfo {author} {\bibfnamefont {M.}~\bibnamefont {Sheindlin}},
  \bibinfo {author} {\bibfnamefont {D.}~\bibnamefont {Staicu}},\ and\ \bibinfo
  {author} {\bibfnamefont {M.}~\bibnamefont {Kinoshita}},\ }\bibfield  {title}
  {\bibinfo {title} {Effect of burn-up on the thermal conductivity of uranium
  dioxide up to 100000mwdt-1},\ }\href
  {https://doi.org/https://doi.org/10.1016/j.jnucmat.2004.01.018} {\bibfield
  {journal} {\bibinfo  {journal} {Journal of Nuclear Materials}\ }\textbf
  {\bibinfo {volume} {327}},\ \bibinfo {pages} {58} (\bibinfo {year}
  {2004})}\BibitemShut {NoStop}%
\bibitem [{\citenamefont {Hua}\ \emph {et~al.}(2024)\citenamefont {Hua},
  \citenamefont {Adnan}, \citenamefont {Khanolkar}, \citenamefont {Rickert},
  \citenamefont {Turner}, \citenamefont {Prusnick}, \citenamefont {Mann},
  \citenamefont {Hurley}, \citenamefont {Khafizov},\ and\ \citenamefont
  {Dennett}}]{HUA_2024709}%
  \BibitemOpen
  \bibfield  {author} {\bibinfo {author} {\bibfnamefont {Z.}~\bibnamefont
  {Hua}}, \bibinfo {author} {\bibfnamefont {S.}~\bibnamefont {Adnan}}, \bibinfo
  {author} {\bibfnamefont {A.~R.}\ \bibnamefont {Khanolkar}}, \bibinfo {author}
  {\bibfnamefont {K.}~\bibnamefont {Rickert}}, \bibinfo {author} {\bibfnamefont
  {D.~B.}\ \bibnamefont {Turner}}, \bibinfo {author} {\bibfnamefont {T.~A.}\
  \bibnamefont {Prusnick}}, \bibinfo {author} {\bibfnamefont {J.~M.}\
  \bibnamefont {Mann}}, \bibinfo {author} {\bibfnamefont {D.~H.}\ \bibnamefont
  {Hurley}}, \bibinfo {author} {\bibfnamefont {M.}~\bibnamefont {Khafizov}},\
  and\ \bibinfo {author} {\bibfnamefont {C.~A.}\ \bibnamefont {Dennett}},\
  }\bibfield  {title} {\bibinfo {title} {Thermal conductivity suppression in
  uranium-doped thorium dioxide due to phonon-spin interactions},\ }\href
  {https://doi.org/https://doi.org/10.1016/j.jmat.2023.11.007} {\bibfield
  {journal} {\bibinfo  {journal} {Journal of Materiomics}\ }\textbf {\bibinfo
  {volume} {10}},\ \bibinfo {pages} {709} (\bibinfo {year} {2024})}\BibitemShut
  {NoStop}%
\bibitem [{\citenamefont {Malakkal}\ \emph {et~al.}(2024)\citenamefont
  {Malakkal}, \citenamefont {Katre}, \citenamefont {Zhou}, \citenamefont
  {Jiang}, \citenamefont {Hurley}, \citenamefont {Marianetti},\ and\
  \citenamefont {Khafizov}}]{Malakkal_PRM}%
  \BibitemOpen
  \bibfield  {author} {\bibinfo {author} {\bibfnamefont {L.}~\bibnamefont
  {Malakkal}}, \bibinfo {author} {\bibfnamefont {A.}~\bibnamefont {Katre}},
  \bibinfo {author} {\bibfnamefont {S.}~\bibnamefont {Zhou}}, \bibinfo {author}
  {\bibfnamefont {C.}~\bibnamefont {Jiang}}, \bibinfo {author} {\bibfnamefont
  {D.~H.}\ \bibnamefont {Hurley}}, \bibinfo {author} {\bibfnamefont {C.~A.}\
  \bibnamefont {Marianetti}},\ and\ \bibinfo {author} {\bibfnamefont
  {M.}~\bibnamefont {Khafizov}},\ }\bibfield  {title} {\bibinfo {title}
  {First-principles determination of the phonon-point defect scattering and
  thermal transport due to fission products in {T}h{O}$_2$},\ }\href
  {https://doi.org/10.1103/PhysRevMaterials.8.025401} {\bibfield  {journal}
  {\bibinfo  {journal} {Phys. Rev. Mater.}\ }\textbf {\bibinfo {volume} {8}},\
  \bibinfo {pages} {025401} (\bibinfo {year} {2024})}\BibitemShut {NoStop}%
\bibitem [{\citenamefont {Kato}\ \emph
  {et~al.}(2024{\natexlab{a}})\citenamefont {Kato}, \citenamefont {Oki},
  \citenamefont {Watanabe}, \citenamefont {Hirooka}, \citenamefont {Vauchy},
  \citenamefont {Ozawa}, \citenamefont {Uwaba}, \citenamefont {Ikusawa},
  \citenamefont {Nakamura},\ and\ \citenamefont {Machida}}]{Masato_2024}%
  \BibitemOpen
  \bibfield  {author} {\bibinfo {author} {\bibfnamefont {M.}~\bibnamefont
  {Kato}}, \bibinfo {author} {\bibfnamefont {T.}~\bibnamefont {Oki}}, \bibinfo
  {author} {\bibfnamefont {M.}~\bibnamefont {Watanabe}}, \bibinfo {author}
  {\bibfnamefont {S.}~\bibnamefont {Hirooka}}, \bibinfo {author} {\bibfnamefont
  {R.}~\bibnamefont {Vauchy}}, \bibinfo {author} {\bibfnamefont
  {T.}~\bibnamefont {Ozawa}}, \bibinfo {author} {\bibfnamefont
  {T.}~\bibnamefont {Uwaba}}, \bibinfo {author} {\bibfnamefont
  {Y.}~\bibnamefont {Ikusawa}}, \bibinfo {author} {\bibfnamefont
  {H.}~\bibnamefont {Nakamura}},\ and\ \bibinfo {author} {\bibfnamefont
  {M.}~\bibnamefont {Machida}},\ }\bibfield  {title} {\bibinfo {title} {A
  science-based mixed oxide property model for developing advanced oxide
  nuclear fuels},\ }\href@noop {} {\bibfield  {journal} {\bibinfo  {journal}
  {Journal of the American Ceramic Society}\ }\textbf {\bibinfo {volume}
  {107}},\ \bibinfo {pages} {2998} (\bibinfo {year}
  {2024}{\natexlab{a}})}\BibitemShut {NoStop}%
\bibitem [{\citenamefont {Saoudi}\ \emph {et~al.}(2018)\citenamefont {Saoudi},
  \citenamefont {Staicu}, \citenamefont {Mouris}, \citenamefont {Bergeron},
  \citenamefont {Hamilton}, \citenamefont {Naji}, \citenamefont {Freis},\ and\
  \citenamefont {Cologna}}]{SAOUDI_2018}%
  \BibitemOpen
  \bibfield  {author} {\bibinfo {author} {\bibfnamefont {M.}~\bibnamefont
  {Saoudi}}, \bibinfo {author} {\bibfnamefont {D.}~\bibnamefont {Staicu}},
  \bibinfo {author} {\bibfnamefont {J.}~\bibnamefont {Mouris}}, \bibinfo
  {author} {\bibfnamefont {A.}~\bibnamefont {Bergeron}}, \bibinfo {author}
  {\bibfnamefont {H.}~\bibnamefont {Hamilton}}, \bibinfo {author}
  {\bibfnamefont {M.}~\bibnamefont {Naji}}, \bibinfo {author} {\bibfnamefont
  {D.}~\bibnamefont {Freis}},\ and\ \bibinfo {author} {\bibfnamefont
  {M.}~\bibnamefont {Cologna}},\ }\bibfield  {title} {\bibinfo {title} {Thermal
  diffusivity and conductivity of thorium- uranium mixed oxides},\ }\href
  {https://doi.org/https://doi.org/10.1016/j.jnucmat.2018.01.014} {\bibfield
  {journal} {\bibinfo  {journal} {Journal of Nuclear Materials}\ }\textbf
  {\bibinfo {volume} {500}},\ \bibinfo {pages} {381} (\bibinfo {year}
  {2018})}\BibitemShut {NoStop}%
\bibitem [{\citenamefont {Turnbull}\ \emph {et~al.}(2023)\citenamefont
  {Turnbull}, \citenamefont {Walker}, \citenamefont {Staicu}, \citenamefont
  {Papaioannou},\ and\ \citenamefont {Yagnik}}]{TURNBULL_2023}%
  \BibitemOpen
  \bibfield  {author} {\bibinfo {author} {\bibfnamefont {J.}~\bibnamefont
  {Turnbull}}, \bibinfo {author} {\bibfnamefont {C.}~\bibnamefont {Walker}},
  \bibinfo {author} {\bibfnamefont {D.}~\bibnamefont {Staicu}}, \bibinfo
  {author} {\bibfnamefont {D.}~\bibnamefont {Papaioannou}},\ and\ \bibinfo
  {author} {\bibfnamefont {S.}~\bibnamefont {Yagnik}},\ }\bibfield  {title}
  {\bibinfo {title} {Effect of burn-up on the thermal conductivity of light
  water reactor fuel: Results of investigations employing the laser flash
  technique},\ }\href
  {https://doi.org/https://doi.org/10.1016/j.jnucmat.2023.154398} {\bibfield
  {journal} {\bibinfo  {journal} {Journal of Nuclear Materials}\ }\textbf
  {\bibinfo {volume} {580}},\ \bibinfo {pages} {154398} (\bibinfo {year}
  {2023})}\BibitemShut {NoStop}%
\bibitem [{\citenamefont {Cooper}\ \emph {et~al.}(2015)\citenamefont {Cooper},
  \citenamefont {Middleburgh},\ and\ \citenamefont {Grimes}}]{COOPER201529}%
  \BibitemOpen
  \bibfield  {author} {\bibinfo {author} {\bibfnamefont {M.}~\bibnamefont
  {Cooper}}, \bibinfo {author} {\bibfnamefont {S.}~\bibnamefont
  {Middleburgh}},\ and\ \bibinfo {author} {\bibfnamefont {R.}~\bibnamefont
  {Grimes}},\ }\bibfield  {title} {\bibinfo {title} {Modelling the thermal
  conductivity of \uppercase{U}$_x$\uppercase{T}h$_{1-x}$\uppercase{O}$_2$ and
  \uppercase{U}$_x$\uppercase{P}u$_{1-x}$\uppercase{O}$_2$},\ }\href
  {https://doi.org/https://doi.org/10.1016/j.jnucmat.2015.07.022} {\bibfield
  {journal} {\bibinfo  {journal} {Journal of Nuclear Materials}\ }\textbf
  {\bibinfo {volume} {466}},\ \bibinfo {pages} {29} (\bibinfo {year}
  {2015})}\BibitemShut {NoStop}%
\bibitem [{\citenamefont {Park}\ \emph {et~al.}(2018)\citenamefont {Park},
  \citenamefont {Farfán}, \citenamefont {Mitchell}, \citenamefont {Resnick},
  \citenamefont {Enriquez},\ and\ \citenamefont {Yee}}]{PARK2018198}%
  \BibitemOpen
  \bibfield  {author} {\bibinfo {author} {\bibfnamefont {J.}~\bibnamefont
  {Park}}, \bibinfo {author} {\bibfnamefont {E.~B.}\ \bibnamefont {Farfán}},
  \bibinfo {author} {\bibfnamefont {K.}~\bibnamefont {Mitchell}}, \bibinfo
  {author} {\bibfnamefont {A.}~\bibnamefont {Resnick}}, \bibinfo {author}
  {\bibfnamefont {C.}~\bibnamefont {Enriquez}},\ and\ \bibinfo {author}
  {\bibfnamefont {T.}~\bibnamefont {Yee}},\ }\bibfield  {title} {\bibinfo
  {title} {Sensitivity of thermal transport in thorium dioxide to defects},\
  }\href {https://doi.org/https://doi.org/10.1016/j.jnucmat.2018.03.043}
  {\bibfield  {journal} {\bibinfo  {journal} {Journal of Nuclear Materials}\
  }\textbf {\bibinfo {volume} {504}},\ \bibinfo {pages} {198} (\bibinfo {year}
  {2018})}\BibitemShut {NoStop}%
\bibitem [{\citenamefont {Rahman}\ \emph {et~al.}(2020)\citenamefont {Rahman},
  \citenamefont {Szpunar},\ and\ \citenamefont {Szpunar}}]{RAHMAN2020152050}%
  \BibitemOpen
  \bibfield  {author} {\bibinfo {author} {\bibfnamefont {M.}~\bibnamefont
  {Rahman}}, \bibinfo {author} {\bibfnamefont {B.}~\bibnamefont {Szpunar}},\
  and\ \bibinfo {author} {\bibfnamefont {J.}~\bibnamefont {Szpunar}},\
  }\bibfield  {title} {\bibinfo {title} {Dependence of thermal conductivity on
  fission-product defects and vacancy concentration in thorium dioxide},\
  }\href {https://doi.org/https://doi.org/10.1016/j.jnucmat.2020.152050}
  {\bibfield  {journal} {\bibinfo  {journal} {Journal of Nuclear Materials}\
  }\textbf {\bibinfo {volume} {532}},\ \bibinfo {pages} {152050} (\bibinfo
  {year} {2020})}\BibitemShut {NoStop}%
\bibitem [{\citenamefont {Jin}\ \emph {et~al.}(2022{\natexlab{b}})\citenamefont
  {Jin}, \citenamefont {Dennett}, \citenamefont {Hurley},\ and\ \citenamefont
  {Khafizov}}]{JIN2022Impact}%
  \BibitemOpen
  \bibfield  {author} {\bibinfo {author} {\bibfnamefont {M.}~\bibnamefont
  {Jin}}, \bibinfo {author} {\bibfnamefont {C.~A.}\ \bibnamefont {Dennett}},
  \bibinfo {author} {\bibfnamefont {D.~H.}\ \bibnamefont {Hurley}},\ and\
  \bibinfo {author} {\bibfnamefont {M.}~\bibnamefont {Khafizov}},\ }\bibfield
  {title} {\bibinfo {title} {Impact of small defects and dislocation loops on
  phonon scattering and thermal transport in \uppercase{T}h\uppercase{O}$_2$},\
  }\href {https://doi.org/https://doi.org/10.1016/j.jnucmat.2022.153758}
  {\bibfield  {journal} {\bibinfo  {journal} {Journal of Nuclear Materials}\ ,\
  \bibinfo {pages} {153758}} (\bibinfo {year}
  {2022}{\natexlab{b}})}\BibitemShut {NoStop}%
\bibitem [{\citenamefont {Deskins}\ \emph {et~al.}(2021)\citenamefont
  {Deskins}, \citenamefont {Hamed}, \citenamefont {Kumagai}, \citenamefont
  {Dennett}, \citenamefont {Peng}, \citenamefont {Khafizov}, \citenamefont
  {Hurley},\ and\ \citenamefont {El-Azab}}]{Deskins_2021}%
  \BibitemOpen
  \bibfield  {author} {\bibinfo {author} {\bibfnamefont {W.~R.}\ \bibnamefont
  {Deskins}}, \bibinfo {author} {\bibfnamefont {A.}~\bibnamefont {Hamed}},
  \bibinfo {author} {\bibfnamefont {T.}~\bibnamefont {Kumagai}}, \bibinfo
  {author} {\bibfnamefont {C.~A.}\ \bibnamefont {Dennett}}, \bibinfo {author}
  {\bibfnamefont {J.}~\bibnamefont {Peng}}, \bibinfo {author} {\bibfnamefont
  {M.}~\bibnamefont {Khafizov}}, \bibinfo {author} {\bibfnamefont
  {D.}~\bibnamefont {Hurley}},\ and\ \bibinfo {author} {\bibfnamefont
  {A.}~\bibnamefont {El-Azab}},\ }\bibfield  {title} {\bibinfo {title} {Thermal
  conductivity of \uppercase{T}h\uppercase{O}$_2$: Effect of point defect
  disorder},\ }\href {https://doi.org/10.1063/5.0038117} {\bibfield  {journal}
  {\bibinfo  {journal} {Journal of Applied Physics}\ }\textbf {\bibinfo
  {volume} {129}},\ \bibinfo {pages} {075102} (\bibinfo {year}
  {2021})}\BibitemShut {NoStop}%
\bibitem [{\citenamefont {Zhou}\ \emph {et~al.}(2018)\citenamefont {Zhou},
  \citenamefont {Fan}, \citenamefont {Qin}, \citenamefont {Yang}, \citenamefont
  {Ouyang},\ and\ \citenamefont {Hu}}]{Zhou_2018}%
  \BibitemOpen
  \bibfield  {author} {\bibinfo {author} {\bibfnamefont {Y.}~\bibnamefont
  {Zhou}}, \bibinfo {author} {\bibfnamefont {Z.}~\bibnamefont {Fan}}, \bibinfo
  {author} {\bibfnamefont {G.}~\bibnamefont {Qin}}, \bibinfo {author}
  {\bibfnamefont {J.-Y.}\ \bibnamefont {Yang}}, \bibinfo {author}
  {\bibfnamefont {T.}~\bibnamefont {Ouyang}},\ and\ \bibinfo {author}
  {\bibfnamefont {M.}~\bibnamefont {Hu}},\ }\bibfield  {title} {\bibinfo
  {title} {Methodology perspective of computing thermal transport in
  low-dimensional materials and nanostructures: The old and the new},\ }\href
  {https://doi.org/10.1021/acsomega.7b01594} {\bibfield  {journal} {\bibinfo
  {journal} {ACS Omega}\ }\textbf {\bibinfo {volume} {3}},\ \bibinfo {pages}
  {3278} (\bibinfo {year} {2018})}\BibitemShut {NoStop}%
\bibitem [{\citenamefont {Jin}\ \emph {et~al.}(2021)\citenamefont {Jin},
  \citenamefont {Khafizov}, \citenamefont {Jiang}, \citenamefont {Zhou},
  \citenamefont {Marianetti}, \citenamefont {Bryan}, \citenamefont {Manley},\
  and\ \citenamefont {Hurley}}]{Jin_2021}%
  \BibitemOpen
  \bibfield  {author} {\bibinfo {author} {\bibfnamefont {M.}~\bibnamefont
  {Jin}}, \bibinfo {author} {\bibfnamefont {M.}~\bibnamefont {Khafizov}},
  \bibinfo {author} {\bibfnamefont {C.}~\bibnamefont {Jiang}}, \bibinfo
  {author} {\bibfnamefont {S.}~\bibnamefont {Zhou}}, \bibinfo {author}
  {\bibfnamefont {C.~A.}\ \bibnamefont {Marianetti}}, \bibinfo {author}
  {\bibfnamefont {M.~S.}\ \bibnamefont {Bryan}}, \bibinfo {author}
  {\bibfnamefont {M.~E.}\ \bibnamefont {Manley}},\ and\ \bibinfo {author}
  {\bibfnamefont {D.~H.}\ \bibnamefont {Hurley}},\ }\bibfield  {title}
  {\bibinfo {title} {Assessment of empirical interatomic potential to predict
  thermal conductivity in \uppercase{T}h\uppercase{O}$_2$ and
  \uppercase{U}\uppercase{O}$_2$},\ }\href
  {https://doi.org/10.1088/1361-648x/abdc8f} {\bibfield  {journal} {\bibinfo
  {journal} {Journal of Physics: Condensed Matter}\ }\textbf {\bibinfo {volume}
  {33}},\ \bibinfo {pages} {275402} (\bibinfo {year} {2021})}\BibitemShut
  {NoStop}%
\bibitem [{\citenamefont {Mann}\ \emph {et~al.}(2010)\citenamefont {Mann},
  \citenamefont {Thompson}, \citenamefont {Serivalsatit}, \citenamefont
  {Tritt}, \citenamefont {Ballato},\ and\ \citenamefont {Kolis}}]{Mann_2010}%
  \BibitemOpen
  \bibfield  {author} {\bibinfo {author} {\bibfnamefont {M.}~\bibnamefont
  {Mann}}, \bibinfo {author} {\bibfnamefont {D.}~\bibnamefont {Thompson}},
  \bibinfo {author} {\bibfnamefont {K.}~\bibnamefont {Serivalsatit}}, \bibinfo
  {author} {\bibfnamefont {T.~M.}\ \bibnamefont {Tritt}}, \bibinfo {author}
  {\bibfnamefont {J.}~\bibnamefont {Ballato}},\ and\ \bibinfo {author}
  {\bibfnamefont {J.}~\bibnamefont {Kolis}},\ }\bibfield  {title} {\bibinfo
  {title} {Hydrothermal growth and thermal property characterization of
  \uppercase{T}h\uppercase{O}$_2$ single crystals},\ }\href
  {https://doi.org/10.1021/cg901308f} {\bibfield  {journal} {\bibinfo
  {journal} {Crystal Growth \& Design}\ }\textbf {\bibinfo {volume} {10}},\
  \bibinfo {pages} {2146} (\bibinfo {year} {2010})}\BibitemShut {NoStop}%
\bibitem [{\citenamefont {Feser}\ and\ \citenamefont
  {Cahill}(2012)}]{Feser_2012}%
  \BibitemOpen
  \bibfield  {author} {\bibinfo {author} {\bibfnamefont {J.~P.}\ \bibnamefont
  {Feser}}\ and\ \bibinfo {author} {\bibfnamefont {D.~G.}\ \bibnamefont
  {Cahill}},\ }\bibfield  {title} {\bibinfo {title} {{Probing anisotropic heat
  transport using time-domain thermoreflectance with offset laser spots}},\
  }\href {https://doi.org/10.1063/1.4757863} {\bibfield  {journal} {\bibinfo
  {journal} {Review of Scientific Instruments}\ }\textbf {\bibinfo {volume}
  {83}},\ \bibinfo {pages} {104901} (\bibinfo {year} {2012})}\BibitemShut
  {NoStop}%
\bibitem [{\citenamefont {Hurley}\ \emph {et~al.}(2015)\citenamefont {Hurley},
  \citenamefont {Schley}, \citenamefont {Khafizov},\ and\ \citenamefont
  {Wendt}}]{Hurley_2015}%
  \BibitemOpen
  \bibfield  {author} {\bibinfo {author} {\bibfnamefont {D.~H.}\ \bibnamefont
  {Hurley}}, \bibinfo {author} {\bibfnamefont {R.~S.}\ \bibnamefont {Schley}},
  \bibinfo {author} {\bibfnamefont {M.}~\bibnamefont {Khafizov}},\ and\
  \bibinfo {author} {\bibfnamefont {B.~L.}\ \bibnamefont {Wendt}},\ }\bibfield
  {title} {\bibinfo {title} {{Local measurement of thermal conductivity and
  diffusivity}},\ }\href {https://doi.org/10.1063/1.4936213} {\bibfield
  {journal} {\bibinfo  {journal} {Review of Scientific Instruments}\ }\textbf
  {\bibinfo {volume} {86}},\ \bibinfo {pages} {123901} (\bibinfo {year}
  {2015})}\BibitemShut {NoStop}%
\bibitem [{\citenamefont {Hua}\ \emph {et~al.}(2012)\citenamefont {Hua},
  \citenamefont {Ban}, \citenamefont {Khafizov}, \citenamefont {Schley},
  \citenamefont {Kennedy},\ and\ \citenamefont {Hurley}}]{Hua_2012}%
  \BibitemOpen
  \bibfield  {author} {\bibinfo {author} {\bibfnamefont {Z.}~\bibnamefont
  {Hua}}, \bibinfo {author} {\bibfnamefont {H.}~\bibnamefont {Ban}}, \bibinfo
  {author} {\bibfnamefont {M.}~\bibnamefont {Khafizov}}, \bibinfo {author}
  {\bibfnamefont {R.}~\bibnamefont {Schley}}, \bibinfo {author} {\bibfnamefont
  {R.}~\bibnamefont {Kennedy}},\ and\ \bibinfo {author} {\bibfnamefont {D.~H.}\
  \bibnamefont {Hurley}},\ }\bibfield  {title} {\bibinfo {title} {{Spatially
  localized measurement of thermal conductivity using a hybrid photothermal
  technique}},\ }\href {https://doi.org/10.1063/1.4716474} {\bibfield
  {journal} {\bibinfo  {journal} {Journal of Applied Physics}\ }\textbf
  {\bibinfo {volume} {111}},\ \bibinfo {pages} {103505} (\bibinfo {year}
  {2012})}\BibitemShut {NoStop}%
\bibitem [{\citenamefont {Maznev}\ \emph {et~al.}(1995)\citenamefont {Maznev},
  \citenamefont {Hartmann},\ and\ \citenamefont {Reichling}}]{Maznev_1995}%
  \BibitemOpen
  \bibfield  {author} {\bibinfo {author} {\bibfnamefont {A.~A.}\ \bibnamefont
  {Maznev}}, \bibinfo {author} {\bibfnamefont {J.}~\bibnamefont {Hartmann}},\
  and\ \bibinfo {author} {\bibfnamefont {M.}~\bibnamefont {Reichling}},\
  }\bibfield  {title} {\bibinfo {title} {{Thermal wave propagation in thin
  films on substrates}},\ }\href {https://doi.org/10.1063/1.359702} {\bibfield
  {journal} {\bibinfo  {journal} {Journal of Applied Physics}\ }\textbf
  {\bibinfo {volume} {78}},\ \bibinfo {pages} {5266} (\bibinfo {year}
  {1995})}\BibitemShut {NoStop}%
\bibitem [{\citenamefont {Wilson}\ \emph {et~al.}(2012)\citenamefont {Wilson},
  \citenamefont {Apgar}, \citenamefont {Martin},\ and\ \citenamefont
  {Cahill}}]{Wilson_2012}%
  \BibitemOpen
  \bibfield  {author} {\bibinfo {author} {\bibfnamefont {R.~B.}\ \bibnamefont
  {Wilson}}, \bibinfo {author} {\bibfnamefont {B.~A.}\ \bibnamefont {Apgar}},
  \bibinfo {author} {\bibfnamefont {L.~W.}\ \bibnamefont {Martin}},\ and\
  \bibinfo {author} {\bibfnamefont {D.~G.}\ \bibnamefont {Cahill}},\ }\bibfield
   {title} {\bibinfo {title} {Thermoreflectance of metal transducers for
  optical pump-probe studies of thermal properties},\ }\href
  {https://doi.org/10.1364/OE.20.028829} {\bibfield  {journal} {\bibinfo
  {journal} {Opt. Express}\ }\textbf {\bibinfo {volume} {20}},\ \bibinfo
  {pages} {28829} (\bibinfo {year} {2012})}\BibitemShut {NoStop}%
\bibitem [{\citenamefont {Kohn}\ and\ \citenamefont {Sham}(1965)}]{Kohn_1965}%
  \BibitemOpen
  \bibfield  {author} {\bibinfo {author} {\bibfnamefont {W.}~\bibnamefont
  {Kohn}}\ and\ \bibinfo {author} {\bibfnamefont {L.~J.}\ \bibnamefont
  {Sham}},\ }\bibfield  {title} {\bibinfo {title} {Self-consistent equations
  including exchange and correlation effects},\ }\href
  {https://doi.org/10.1103/PhysRev.140.A1133} {\bibfield  {journal} {\bibinfo
  {journal} {Phys. Rev.}\ }\textbf {\bibinfo {volume} {140}},\ \bibinfo {pages}
  {A1133} (\bibinfo {year} {1965})}\BibitemShut {NoStop}%
\bibitem [{\citenamefont {Bl\"ochl}(1994)}]{blochl_1994}%
  \BibitemOpen
  \bibfield  {author} {\bibinfo {author} {\bibfnamefont {P.~E.}\ \bibnamefont
  {Bl\"ochl}},\ }\bibfield  {title} {\bibinfo {title} {Projector augmented-wave
  method},\ }\href {https://doi.org/10.1103/PhysRevB.50.17953} {\bibfield
  {journal} {\bibinfo  {journal} {Phys. Rev. B}\ }\textbf {\bibinfo {volume}
  {50}},\ \bibinfo {pages} {17953} (\bibinfo {year} {1994})}\BibitemShut
  {NoStop}%
\bibitem [{\citenamefont {Kresse}\ and\ \citenamefont
  {Furthm\"uller}(1996)}]{Kresse_1996}%
  \BibitemOpen
  \bibfield  {author} {\bibinfo {author} {\bibfnamefont {G.}~\bibnamefont
  {Kresse}}\ and\ \bibinfo {author} {\bibfnamefont {J.}~\bibnamefont
  {Furthm\"uller}},\ }\bibfield  {title} {\bibinfo {title} {Efficient iterative
  schemes for ab initio total-energy calculations using a plane-wave basis
  set},\ }\href {https://doi.org/10.1103/PhysRevB.54.11169} {\bibfield
  {journal} {\bibinfo  {journal} {Phys. Rev. B}\ }\textbf {\bibinfo {volume}
  {54}},\ \bibinfo {pages} {11169} (\bibinfo {year} {1996})}\BibitemShut
  {NoStop}%
\bibitem [{\citenamefont {Perdew}\ and\ \citenamefont
  {Zunger}(1981)}]{Perdew_1981}%
  \BibitemOpen
  \bibfield  {author} {\bibinfo {author} {\bibfnamefont {J.~P.}\ \bibnamefont
  {Perdew}}\ and\ \bibinfo {author} {\bibfnamefont {A.}~\bibnamefont
  {Zunger}},\ }\bibfield  {title} {\bibinfo {title} {Self-interaction
  correction to density-functional approximations for many-electron systems},\
  }\href {https://doi.org/10.1103/PhysRevB.23.5048} {\bibfield  {journal}
  {\bibinfo  {journal} {Phys. Rev. B}\ }\textbf {\bibinfo {volume} {23}},\
  \bibinfo {pages} {5048} (\bibinfo {year} {1981})}\BibitemShut {NoStop}%
\bibitem [{\citenamefont {Dudarev}\ \emph {et~al.}(1998)\citenamefont
  {Dudarev}, \citenamefont {Botton}, \citenamefont {Savrasov}, \citenamefont
  {Humphreys},\ and\ \citenamefont {Sutton}}]{Dudarev_1988_HubbardU}%
  \BibitemOpen
  \bibfield  {author} {\bibinfo {author} {\bibfnamefont {S.~L.}\ \bibnamefont
  {Dudarev}}, \bibinfo {author} {\bibfnamefont {G.~A.}\ \bibnamefont {Botton}},
  \bibinfo {author} {\bibfnamefont {S.~Y.}\ \bibnamefont {Savrasov}}, \bibinfo
  {author} {\bibfnamefont {C.~J.}\ \bibnamefont {Humphreys}},\ and\ \bibinfo
  {author} {\bibfnamefont {A.~P.}\ \bibnamefont {Sutton}},\ }\bibfield  {title}
  {\bibinfo {title} {Electron-energy-loss spectra and the structural stability
  of nickel oxide: An {LSDA+U} study},\ }\href
  {https://doi.org/10.1103/PhysRevB.57.1505} {\bibfield  {journal} {\bibinfo
  {journal} {Phys. Rev. B}\ }\textbf {\bibinfo {volume} {57}},\ \bibinfo
  {pages} {1505} (\bibinfo {year} {1998})}\BibitemShut {NoStop}%
\bibitem [{\citenamefont {Shields}\ \emph {et~al.}(2016)\citenamefont
  {Shields}, \citenamefont {Santos-Carballal},\ and\ \citenamefont {{de
  Leeuw}}}]{SHIELDS201699}%
  \BibitemOpen
  \bibfield  {author} {\bibinfo {author} {\bibfnamefont {A.~E.}\ \bibnamefont
  {Shields}}, \bibinfo {author} {\bibfnamefont {D.}~\bibnamefont
  {Santos-Carballal}},\ and\ \bibinfo {author} {\bibfnamefont {N.~H.}\
  \bibnamefont {{de Leeuw}}},\ }\bibfield  {title} {\bibinfo {title} {A density
  functional theory study of uranium-doped thoria and uranium adatoms on the
  major surfaces of thorium dioxide},\ }\href
  {https://doi.org/https://doi.org/10.1016/j.jnucmat.2016.02.009} {\bibfield
  {journal} {\bibinfo  {journal} {Journal of Nuclear Materials}\ }\textbf
  {\bibinfo {volume} {473}},\ \bibinfo {pages} {99} (\bibinfo {year}
  {2016})}\BibitemShut {NoStop}%
\bibitem [{\citenamefont {Lu}\ \emph {et~al.}(2012)\citenamefont {Lu},
  \citenamefont {Yang},\ and\ \citenamefont {Zhang}}]{Lu_2012}%
  \BibitemOpen
  \bibfield  {author} {\bibinfo {author} {\bibfnamefont {Y.}~\bibnamefont
  {Lu}}, \bibinfo {author} {\bibfnamefont {Y.}~\bibnamefont {Yang}},\ and\
  \bibinfo {author} {\bibfnamefont {P.}~\bibnamefont {Zhang}},\ }\bibfield
  {title} {\bibinfo {title} {Thermodynamic properties and structural stability
  of thorium dioxide},\ }\href {https://doi.org/10.1088/0953-8984/24/22/225801}
  {\bibfield  {journal} {\bibinfo  {journal} {Journal of Physics: Condensed
  Matter}\ }\textbf {\bibinfo {volume} {24}},\ \bibinfo {pages} {225801}
  (\bibinfo {year} {2012})}\BibitemShut {NoStop}%
\bibitem [{\citenamefont {Togo}\ and\ \citenamefont {Tanaka}(2015)}]{TOGO2015}%
  \BibitemOpen
  \bibfield  {author} {\bibinfo {author} {\bibfnamefont {A.}~\bibnamefont
  {Togo}}\ and\ \bibinfo {author} {\bibfnamefont {I.}~\bibnamefont {Tanaka}},\
  }\bibfield  {title} {\bibinfo {title} {First principles phonon calculations
  in materials science},\ }\href
  {https://doi.org/https://doi.org/10.1016/j.scriptamat.2015.07.021} {\bibfield
   {journal} {\bibinfo  {journal} {Scripta Materialia}\ }\textbf {\bibinfo
  {volume} {108}},\ \bibinfo {pages} {1} (\bibinfo {year} {2015})}\BibitemShut
  {NoStop}%
\bibitem [{\citenamefont {Li}\ \emph {et~al.}(2014)\citenamefont {Li},
  \citenamefont {Carrete}, \citenamefont {{A. Katcho}},\ and\ \citenamefont
  {Mingo}}]{shengBTE_2014}%
  \BibitemOpen
  \bibfield  {author} {\bibinfo {author} {\bibfnamefont {W.}~\bibnamefont
  {Li}}, \bibinfo {author} {\bibfnamefont {J.}~\bibnamefont {Carrete}},
  \bibinfo {author} {\bibfnamefont {N.}~\bibnamefont {{A. Katcho}}},\ and\
  \bibinfo {author} {\bibfnamefont {N.}~\bibnamefont {Mingo}},\ }\bibfield
  {title} {\bibinfo {title} {Shengbte: A solver of the boltzmann transport
  equation for phonons},\ }\href
  {https://doi.org/https://doi.org/10.1016/j.cpc.2014.02.015} {\bibfield
  {journal} {\bibinfo  {journal} {Computer Physics Communications}\ }\textbf
  {\bibinfo {volume} {185}},\ \bibinfo {pages} {1747} (\bibinfo {year}
  {2014})}\BibitemShut {NoStop}%
\bibitem [{\citenamefont {Yi}\ \emph {et~al.}(2016)\citenamefont {Yi},
  \citenamefont {Shun-Li}, \citenamefont {Huazhi}, \citenamefont {Zi-Kui},\
  and\ \citenamefont {Qing}}]{Wang2016}%
  \BibitemOpen
  \bibfield  {author} {\bibinfo {author} {\bibfnamefont {W.}~\bibnamefont
  {Yi}}, \bibinfo {author} {\bibfnamefont {S.}~\bibnamefont {Shun-Li}},
  \bibinfo {author} {\bibfnamefont {F.}~\bibnamefont {Huazhi}}, \bibinfo
  {author} {\bibfnamefont {L.}~\bibnamefont {Zi-Kui}},\ and\ \bibinfo {author}
  {\bibfnamefont {C.~L.}\ \bibnamefont {Qing}},\ }\bibfield  {title} {\bibinfo
  {title} {First-principles calculations of lattice dynamics and thermal
  properties of polar solids},\ }\href
  {https://doi.org/10.1038/npjcompumats.2016.6} {\bibfield  {journal} {\bibinfo
   {journal} {npj Computational Materials}\ }\textbf {\bibinfo {volume} {2}},\
  \bibinfo {pages} {16006} (\bibinfo {year} {2016})}\BibitemShut {NoStop}%
\bibitem [{\citenamefont {Baroni}\ \emph {et~al.}(2001)\citenamefont {Baroni},
  \citenamefont {De~Gironcoli}, \citenamefont {Dal~Corso},\ and\ \citenamefont
  {Giannozzi}}]{Baroni_2001_Phonons}%
  \BibitemOpen
  \bibfield  {author} {\bibinfo {author} {\bibfnamefont {S.}~\bibnamefont
  {Baroni}}, \bibinfo {author} {\bibfnamefont {S.}~\bibnamefont
  {De~Gironcoli}}, \bibinfo {author} {\bibfnamefont {A.}~\bibnamefont
  {Dal~Corso}},\ and\ \bibinfo {author} {\bibfnamefont {P.}~\bibnamefont
  {Giannozzi}},\ }\bibfield  {title} {\bibinfo {title} {Phonons and related
  crystal properties from density-functional perturbation theory},\ }\href
  {https://doi.org/10.1103/revmodphys.73.515} {\bibfield  {journal} {\bibinfo
  {journal} {Reviews of Modern Physics}\ }\textbf {\bibinfo {volume} {73}},\
  \bibinfo {pages} {515–562} (\bibinfo {year} {2001})}\BibitemShut {NoStop}%
\bibitem [{\citenamefont {Tamura}(1983)}]{Tamura_1983}%
  \BibitemOpen
  \bibfield  {author} {\bibinfo {author} {\bibfnamefont {S.-I.}\ \bibnamefont
  {Tamura}},\ }\bibfield  {title} {\bibinfo {title} {Isotope scattering of
  dispersive phonons in {G}e},\ }\href
  {https://doi.org/10.1103/PhysRevB.27.858} {\bibfield  {journal} {\bibinfo
  {journal} {Phys. Rev. B}\ }\textbf {\bibinfo {volume} {27}},\ \bibinfo
  {pages} {858} (\bibinfo {year} {1983})}\BibitemShut {NoStop}%
\bibitem [{\citenamefont {Katre}\ \emph {et~al.}(2017)\citenamefont {Katre},
  \citenamefont {Carrete}, \citenamefont {Dongre}, \citenamefont {Madsen},\
  and\ \citenamefont {Mingo}}]{Ankita_2017}%
  \BibitemOpen
  \bibfield  {author} {\bibinfo {author} {\bibfnamefont {A.}~\bibnamefont
  {Katre}}, \bibinfo {author} {\bibfnamefont {J.}~\bibnamefont {Carrete}},
  \bibinfo {author} {\bibfnamefont {B.}~\bibnamefont {Dongre}}, \bibinfo
  {author} {\bibfnamefont {G.~K.~H.}\ \bibnamefont {Madsen}},\ and\ \bibinfo
  {author} {\bibfnamefont {N.}~\bibnamefont {Mingo}},\ }\bibfield  {title}
  {\bibinfo {title} {Exceptionally strong phonon scattering by {B} substitution
  in cubic {S}i{C}},\ }\href {https://doi.org/10.1103/PhysRevLett.119.075902}
  {\bibfield  {journal} {\bibinfo  {journal} {Phys. Rev. Lett.}\ }\textbf
  {\bibinfo {volume} {119}},\ \bibinfo {pages} {075902} (\bibinfo {year}
  {2017})}\BibitemShut {NoStop}%
\bibitem [{\citenamefont {Xiao}\ \emph {et~al.}(2022)\citenamefont {Xiao},
  \citenamefont {Ma}, \citenamefont {Bryan}, \citenamefont {Fu}, \citenamefont
  {Mann}, \citenamefont {Winn}, \citenamefont {Abernathy}, \citenamefont
  {Hermann}, \citenamefont {Khanolkar}, \citenamefont {Dennett}, \citenamefont
  {Hurley}, \citenamefont {Manley},\ and\ \citenamefont
  {Marianetti}}]{Enda_2022_prb}%
  \BibitemOpen
  \bibfield  {author} {\bibinfo {author} {\bibfnamefont {E.}~\bibnamefont
  {Xiao}}, \bibinfo {author} {\bibfnamefont {H.}~\bibnamefont {Ma}}, \bibinfo
  {author} {\bibfnamefont {M.~S.}\ \bibnamefont {Bryan}}, \bibinfo {author}
  {\bibfnamefont {L.}~\bibnamefont {Fu}}, \bibinfo {author} {\bibfnamefont
  {J.~M.}\ \bibnamefont {Mann}}, \bibinfo {author} {\bibfnamefont
  {B.}~\bibnamefont {Winn}}, \bibinfo {author} {\bibfnamefont {D.~L.}\
  \bibnamefont {Abernathy}}, \bibinfo {author} {\bibfnamefont {R.~P.}\
  \bibnamefont {Hermann}}, \bibinfo {author} {\bibfnamefont {A.~R.}\
  \bibnamefont {Khanolkar}}, \bibinfo {author} {\bibfnamefont {C.~A.}\
  \bibnamefont {Dennett}}, \bibinfo {author} {\bibfnamefont {D.~H.}\
  \bibnamefont {Hurley}}, \bibinfo {author} {\bibfnamefont {M.~E.}\
  \bibnamefont {Manley}},\ and\ \bibinfo {author} {\bibfnamefont {C.~A.}\
  \bibnamefont {Marianetti}},\ }\bibfield  {title} {\bibinfo {title}
  {Validating first-principles phonon lifetimes via inelastic neutron
  scattering},\ }\href {https://doi.org/10.1103/PhysRevB.106.144310} {\bibfield
   {journal} {\bibinfo  {journal} {Phys. Rev. B}\ }\textbf {\bibinfo {volume}
  {106}},\ \bibinfo {pages} {144310} (\bibinfo {year} {2022})}\BibitemShut
  {NoStop}%
\bibitem [{\citenamefont {Keramidas}\ and\ \citenamefont
  {White}(1973)}]{Keramidas_1973}%
  \BibitemOpen
  \bibfield  {author} {\bibinfo {author} {\bibfnamefont {V.~G.}\ \bibnamefont
  {Keramidas}}\ and\ \bibinfo {author} {\bibfnamefont {W.~B.}\ \bibnamefont
  {White}},\ }\bibfield  {title} {\bibinfo {title} {{Raman spectra of oxides
  with the fluorite structure}},\ }\href {https://doi.org/10.1063/1.1680227}
  {\bibfield  {journal} {\bibinfo  {journal} {The Journal of Chemical Physics}\
  }\textbf {\bibinfo {volume} {59}},\ \bibinfo {pages} {1561} (\bibinfo {year}
  {1973})}\BibitemShut {NoStop}%
\bibitem [{\citenamefont {Rickert}\ \emph {et~al.}(2022)\citenamefont
  {Rickert}, \citenamefont {Prusnick}, \citenamefont {Hunt}, \citenamefont
  {French}, \citenamefont {Turner}, \citenamefont {Dennett}, \citenamefont
  {Shao},\ and\ \citenamefont {Mann}}]{RICKERT_2022}%
  \BibitemOpen
  \bibfield  {author} {\bibinfo {author} {\bibfnamefont {K.}~\bibnamefont
  {Rickert}}, \bibinfo {author} {\bibfnamefont {T.~A.}\ \bibnamefont
  {Prusnick}}, \bibinfo {author} {\bibfnamefont {E.}~\bibnamefont {Hunt}},
  \bibinfo {author} {\bibfnamefont {A.}~\bibnamefont {French}}, \bibinfo
  {author} {\bibfnamefont {D.~B.}\ \bibnamefont {Turner}}, \bibinfo {author}
  {\bibfnamefont {C.~A.}\ \bibnamefont {Dennett}}, \bibinfo {author}
  {\bibfnamefont {L.}~\bibnamefont {Shao}},\ and\ \bibinfo {author}
  {\bibfnamefont {J.~M.}\ \bibnamefont {Mann}},\ }\bibfield  {title} {\bibinfo
  {title} {Raman and photoluminescence evaluation of ion-induced damage
  uniformity in tho2},\ }\href
  {https://doi.org/https://doi.org/10.1016/j.nimb.2022.01.011} {\bibfield
  {journal} {\bibinfo  {journal} {Nuclear Instruments and Methods in Physics
  Research Section B: Beam Interactions with Materials and Atoms}\ }\textbf
  {\bibinfo {volume} {515}},\ \bibinfo {pages} {69} (\bibinfo {year}
  {2022})}\BibitemShut {NoStop}%
\bibitem [{\citenamefont {Mock}\ \emph {et~al.}(2019)\citenamefont {Mock},
  \citenamefont {Dugan}, \citenamefont {Knight}, \citenamefont {Korlacki},
  \citenamefont {Mann}, \citenamefont {Kimani}, \citenamefont {Petrosky},
  \citenamefont {Dowben},\ and\ \citenamefont {Schubert}}]{Mock_2019}%
  \BibitemOpen
  \bibfield  {author} {\bibinfo {author} {\bibfnamefont {A.}~\bibnamefont
  {Mock}}, \bibinfo {author} {\bibfnamefont {C.}~\bibnamefont {Dugan}},
  \bibinfo {author} {\bibfnamefont {S.}~\bibnamefont {Knight}}, \bibinfo
  {author} {\bibfnamefont {R.}~\bibnamefont {Korlacki}}, \bibinfo {author}
  {\bibfnamefont {J.~M.}\ \bibnamefont {Mann}}, \bibinfo {author}
  {\bibfnamefont {M.~M.}\ \bibnamefont {Kimani}}, \bibinfo {author}
  {\bibfnamefont {J.~C.}\ \bibnamefont {Petrosky}}, \bibinfo {author}
  {\bibfnamefont {P.~A.}\ \bibnamefont {Dowben}},\ and\ \bibinfo {author}
  {\bibfnamefont {M.}~\bibnamefont {Schubert}},\ }\bibfield  {title} {\bibinfo
  {title} {{Band-to-band transitions and critical points in the near-infrared
  to vacuum ultraviolet dielectric functions of single crystal urania and
  thoria}},\ }\href {https://doi.org/10.1063/1.5087059} {\bibfield  {journal}
  {\bibinfo  {journal} {Applied Physics Letters}\ }\textbf {\bibinfo {volume}
  {114}},\ \bibinfo {pages} {211901} (\bibinfo {year} {2019})}\BibitemShut
  {NoStop}%
\bibitem [{\citenamefont {Kato}\ \emph
  {et~al.}(2024{\natexlab{b}})\citenamefont {Kato}, \citenamefont {Oki},
  \citenamefont {Watanabe}, \citenamefont {Hirooka}, \citenamefont {Vauchy},
  \citenamefont {Ozawa}, \citenamefont {Uwaba}, \citenamefont {Ikusawa},
  \citenamefont {Nakamura},\ and\ \citenamefont {Machida}}]{kato2024}%
  \BibitemOpen
  \bibfield  {author} {\bibinfo {author} {\bibfnamefont {M.}~\bibnamefont
  {Kato}}, \bibinfo {author} {\bibfnamefont {T.}~\bibnamefont {Oki}}, \bibinfo
  {author} {\bibfnamefont {M.}~\bibnamefont {Watanabe}}, \bibinfo {author}
  {\bibfnamefont {S.}~\bibnamefont {Hirooka}}, \bibinfo {author} {\bibfnamefont
  {R.}~\bibnamefont {Vauchy}}, \bibinfo {author} {\bibfnamefont
  {T.}~\bibnamefont {Ozawa}}, \bibinfo {author} {\bibfnamefont
  {T.}~\bibnamefont {Uwaba}}, \bibinfo {author} {\bibfnamefont
  {Y.}~\bibnamefont {Ikusawa}}, \bibinfo {author} {\bibfnamefont
  {H.}~\bibnamefont {Nakamura}},\ and\ \bibinfo {author} {\bibfnamefont
  {M.}~\bibnamefont {Machida}},\ }\bibfield  {title} {\bibinfo {title} {A
  science-based mixed oxide property model for developing advanced oxide
  nuclear fuels},\ }\href {https://doi.org/https://doi.org/10.1111/jace.19609}
  {\bibfield  {journal} {\bibinfo  {journal} {Journal of the American Ceramic
  Society}\ }\textbf {\bibinfo {volume} {107}},\ \bibinfo {pages} {2998}
  (\bibinfo {year} {2024}{\natexlab{b}})}\BibitemShut {NoStop}%
\bibitem [{\citenamefont {Fukushima}\ \emph {et~al.}(1986)\citenamefont
  {Fukushima}, \citenamefont {Ohmichi},\ and\ \citenamefont
  {Handa}}]{FUKUSHIMA1986}%
  \BibitemOpen
  \bibfield  {author} {\bibinfo {author} {\bibfnamefont {S.}~\bibnamefont
  {Fukushima}}, \bibinfo {author} {\bibfnamefont {T.}~\bibnamefont {Ohmichi}},\
  and\ \bibinfo {author} {\bibfnamefont {M.}~\bibnamefont {Handa}},\ }\bibfield
   {title} {\bibinfo {title} {The effect of rare earths on thermal conductivity
  of uranium, plutonium and their mived ovide fuels},\ }\href
  {https://doi.org/https://doi.org/10.1016/0022-5088(86)90579-5} {\bibfield
  {journal} {\bibinfo  {journal} {Journal of the Less Common Metals}\ }\textbf
  {\bibinfo {volume} {121}},\ \bibinfo {pages} {631} (\bibinfo {year}
  {1986})},\ \bibinfo {note} {proceedings of Actinides 85, Aix en Provence -
  Part I}\BibitemShut {NoStop}%
\bibitem [{\citenamefont {Duriez}\ \emph {et~al.}(2000)\citenamefont {Duriez},
  \citenamefont {Alessandri}, \citenamefont {Gervais},\ and\ \citenamefont
  {Philipponneau}}]{DURIEZ2000}%
  \BibitemOpen
  \bibfield  {author} {\bibinfo {author} {\bibfnamefont {C.}~\bibnamefont
  {Duriez}}, \bibinfo {author} {\bibfnamefont {J.-P.}\ \bibnamefont
  {Alessandri}}, \bibinfo {author} {\bibfnamefont {T.}~\bibnamefont
  {Gervais}},\ and\ \bibinfo {author} {\bibfnamefont {Y.}~\bibnamefont
  {Philipponneau}},\ }\bibfield  {title} {\bibinfo {title} {Thermal
  conductivity of hypostoichiometric low pu content (upu)o2-x mixed oxide},\
  }\href {https://www.sciencedirect.com/science/article/pii/S0022311599002056}
  {\bibfield  {journal} {\bibinfo  {journal} {Journal of Nuclear Materials}\
  }\textbf {\bibinfo {volume} {277}},\ \bibinfo {pages} {143} (\bibinfo {year}
  {2000})}\BibitemShut {NoStop}%
\bibitem [{\citenamefont {Chauhan}\ \emph {et~al.}(2021)\citenamefont
  {Chauhan}, \citenamefont {Pakarinen}, \citenamefont {Yao}, \citenamefont
  {He}, \citenamefont {Hurley},\ and\ \citenamefont {Khafizov}}]{chauhan2021}%
  \BibitemOpen
  \bibfield  {author} {\bibinfo {author} {\bibfnamefont {V.~S.}\ \bibnamefont
  {Chauhan}}, \bibinfo {author} {\bibfnamefont {J.}~\bibnamefont {Pakarinen}},
  \bibinfo {author} {\bibfnamefont {T.}~\bibnamefont {Yao}}, \bibinfo {author}
  {\bibfnamefont {L.}~\bibnamefont {He}}, \bibinfo {author} {\bibfnamefont
  {D.~H.}\ \bibnamefont {Hurley}},\ and\ \bibinfo {author} {\bibfnamefont
  {M.}~\bibnamefont {Khafizov}},\ }\bibfield  {title} {\bibinfo {title}
  {Indirect characterization of point defects in proton irradiated ceria},\
  }\href {https://doi.org/https://doi.org/10.1016/j.mtla.2021.101019}
  {\bibfield  {journal} {\bibinfo  {journal} {Materialia}\ }\textbf {\bibinfo
  {volume} {15}},\ \bibinfo {pages} {101019} (\bibinfo {year}
  {2021})}\BibitemShut {NoStop}%
\bibitem [{\citenamefont {Hua}\ \emph {et~al.}(2023)\citenamefont {Hua},
  \citenamefont {Adnan}, \citenamefont {Khanolkar}, \citenamefont {Rickert},
  \citenamefont {Turner}, \citenamefont {Prusnick}, \citenamefont {Mann},
  \citenamefont {Hurley}, \citenamefont {Khafizov},\ and\ \citenamefont
  {Dennett}}]{hua2023thermal}%
  \BibitemOpen
  \bibfield  {author} {\bibinfo {author} {\bibfnamefont {Z.}~\bibnamefont
  {Hua}}, \bibinfo {author} {\bibfnamefont {S.}~\bibnamefont {Adnan}}, \bibinfo
  {author} {\bibfnamefont {A.~R.}\ \bibnamefont {Khanolkar}}, \bibinfo {author}
  {\bibfnamefont {K.}~\bibnamefont {Rickert}}, \bibinfo {author} {\bibfnamefont
  {D.~B.}\ \bibnamefont {Turner}}, \bibinfo {author} {\bibfnamefont {T.~A.}\
  \bibnamefont {Prusnick}}, \bibinfo {author} {\bibfnamefont {J.~M.}\
  \bibnamefont {Mann}}, \bibinfo {author} {\bibfnamefont {D.~H.}\ \bibnamefont
  {Hurley}}, \bibinfo {author} {\bibfnamefont {M.}~\bibnamefont {Khafizov}},\
  and\ \bibinfo {author} {\bibfnamefont {C.~A.}\ \bibnamefont {Dennett}},\
  }\href@noop {} {\bibinfo {title} {Thermal conductivity suppression in
  uranium-doped thorium dioxide due to phonon resonant scattering}} (\bibinfo
  {year} {2023}),\ \Eprint {https://arxiv.org/abs/2303.01659} {arXiv:2303.01659
  [cond-mat.mtrl-sci]} \BibitemShut {NoStop}%
\bibitem [{\citenamefont {Shannon}(1976)}]{Shannon:a12967}%
  \BibitemOpen
  \bibfield  {author} {\bibinfo {author} {\bibfnamefont {R.~D.}\ \bibnamefont
  {Shannon}},\ }\bibfield  {title} {\bibinfo {title} {{Revised effective ionic
  radii and systematic studies of interatomic distances in halides and
  chalcogenides}},\ }\href {https://doi.org/10.1107/S0567739476001551}
  {\bibfield  {journal} {\bibinfo  {journal} {Acta Crystallographica Section
  A}\ }\textbf {\bibinfo {volume} {32}},\ \bibinfo {pages} {751} (\bibinfo
  {year} {1976})}\BibitemShut {NoStop}%
\end{thebibliography}%

\end{document}